\documentclass[aps,pra,nofootinbib,amsmath,amssymb,showpacs]{revtex4}
\usepackage{graphicx} 
\usepackage{bm}
\usepackage{dcolumn}
\usepackage{amsmath}
\usepackage{epstopdf}
\usepackage{natbib}
\usepackage{subfig}
\usepackage{comment}
\usepackage{engord}
\usepackage{hyperref}
\begin{document}

\title{Density of states and excitonic condensation in the double layer correlated systems}

\author{V. Apinyan\footnote{Corresponding author. Tel.:  +48 71 3954 284; E-mail address: v.apinyan@int.pan.wroc.pl.}, T. K. Kope\'{c}}
\affiliation{Institute for Low Temperature and Structure Research, Polish Academy of Sciences\\
PO. Box 1410, 50-950 Wroc\l{}aw 2, Poland \\}

\date{\today}

\begin{abstract}
%
\textbf{ABSTRACT}

We consider the single-particle density of states (DOS) in the strongly correlated double layer (DL) system, without applied external fields. We demonstrate an unusual collapse effect in the spectrum of the normal single-particle spectral function at the particular high-symmetry point corresponding to the specific bunching-point solution of the chemical potential in the Frenkel channel. We show that at the low-temperature limit the anomalous spectral function obeys a concave like structure, which is directly related to the interlayer pair formation and condensation. We calculate the normal DOS functions, and we find their temperature dependence for different values of the interlayer Coulomb interaction parameter. We show that the normal electron and hole DOS functions demonstrate typical condensates double peak structures on the background of the excitonic pair formation quasiparticle spectra and we have found the evidence of the hybridization gap in the case of high-temperature limit, and small interlayer coupling parameter. Meanwhile, we show a possible crossover from the excitonic condensate regime into the band insulator state. The structure of the normal DOS spectra, in the Frenkel channel and for the strong interlayer coupling regime is found gapless for all temperature limits, which clearly indicates the strong coherence effects in the DL structure, and the excitonic condensates therein. We have shown that the excitonic pair formation and pair condensation occurs simultaneously in the DL system, in contrast with the purely three-dimensional (3D) or two-dimensional cases (2D), discussed previously.  
\end{abstract}

\pacs{71.10.Fd, 71.28.+d, 71.35.Lk, 71.35.-y, 71.10.Hf}
  \maketitle

\renewcommand\thesection{\arabic{section}}
\section{\label{sec:Section_1} Introduction}
%
The physics, related to the electronic double layer (DL) heterostructures, becomes more and more attractive, related to its large applications in the field 
of electronic nanotechnologies \cite{cite-1,cite-2,cite-3,cite-4,cite-5,cite-6,cite-7,cite-8,cite-9}. Thanks to the discovery of the long-lived 100 ns 
coherent electron spin states in n-type semiconductors \cite{cite-10,cite-11,cite-12,cite-13,cite-14,cite-15,cite-16,cite-17}, the study of spintronics in the DL systems has attracted tremendous 
attention in recent years, both in theoretical and 
experimental aspects \cite{cite-18}. The unique properties of energy
dispersion and the density of states (DOS) in spintronic materials
can be used to control and tune the excitonic plasma excitation.

An electron-hole DL system \cite{cite-19, cite-20} is composed of an 
electron layer, and the hole layer, separated by a finite distance. 
In the complex and rich physics related to the electron-hole DL structures, it is important to emphasize the aspects, related to the electron-hole pairing and exciton condensation. To control the formation of the excitonic condensates in the DL structured systems, an external electric, or magnetic field are generally supposed to be applied. The possibility of the excitonic condensates formation in two-dimensions (2D), under the strong magnetic field in a semiconductor heterostructure, with a very field-sensitive band structure, is given in Refs.\onlinecite{cite-21} and \onlinecite{cite-22}. A thermodynamically stable condensate formation in InAs-GaSb -based system is discussed in Ref.\onlinecite{cite-23}, where a study of the Keldysh-BCS theory is given. Meanwhile, a new platform material is proposed for studying the excitonic condensates. Recently, the detailed phase diagram for the inverted InAs/GaSb DL quantum wells has been proposed \cite{cite-24}, and the possibility of the stable $s$-wave exciton condensed phase is proved. Contrary, for the large interlayer tunneling amplitude, a topologically non-trivial quantum spin Hall's insulator phase is obtained, with the $p$-wave pairing gap. 

A smooth crossover from the intralayer Cooper pair superfluid state to interlayer pairing superfluid state is discussed within the extended Hubbard model in Ref.\onlinecite{cite-25} when discussing the possibility to tunnel between the layers in the correlated DL structure. It is shown that the interlayer superfluid state is related to the bilayer exciton condensates formation, where the pairing happens between electrons in one layer and holes in the other. Particularly, it is shown that the crossover smoothening is not significant for the large interlayer hopping and is negligible for the very small interlayer hopping. In the other work \cite{cite-26}, based on the Quantum Monte Carlo (QMC) numerical approach of the stochastic series expansion, the bilayer extended Bose-Hubbard model is studied, and the intralayer and interlayer interactions are considered. A coexistence of the valence bond solid state with the boson superfluidity is found, and the strong interlayer hopping regime is discussed. Meanwhile, it is shown that the valence bond Mott insulator state takes place at the weak repulsive intraplane Coulomb interaction.   

It was shown recently \cite{cite-27, cite-28, cite-29, cite-30, cite-31, cite-32, cite-33}, that for the case of the three-dimensional (3D) bulk semiconductors, the excitonic condensation and the excitonic pair formation are two different phases of matter. Moreover, for the pure two-dimensional electron-hole system, it is shown \cite{cite-30} that the low-temperature vortex-antivortex bounded superfluid phase transition critical temperature differs considerably from the excitonic pair formation critical temperature.    

The DL structure, without applied external field, is
studied in details in Ref.\onlinecite{cite-33}, where a new phase transition
scenario for the excitonic pairing processes is given.
Particularly, a complicated behavior for the excitonic gap
parameter is found, and the existence of both
Frenkel and Wannier-Mott (WM) pairing channels in the gap
structure is shown. It is shown in Ref.\onlinecite{cite-33} that these pairing channels are
continuously merging one into another when tuning the
intralayer Coulomb interaction at the specific merging point.   

In this paper, we consider the DL structure without the external electric or magnetic field, basing on the results given in Ref.\onlinecite{cite-33} and we calculate the single-particle normal electron and hole DOS functions. 
Firstly, basing on the single bunching-point solution in the chemical potential structure of the DL system, we calculate the normal spectral functions in the Frenkel channel and we show an interesting collapse effect in the behavior of the normal spectral functions. Then we calculate the normal DOS spectra. We will show that the normal, electron and hole single-particle DOS spectra are gapless for the low-temperature limit and exhibit a large hybridization gap for the case of high temperatures, and small interlayer Coulomb interaction. Thus, we derive a crossover mechanism, from the condensate state  to the Band Insulator (BI) regime. Regarding the works given in Refs.\onlinecite{cite-27, cite-28, cite-29, cite-30, cite-33}, and as the result of our model, we suggest that in the DL structures, and at the given parameters range, the excitonic condensation and pair formation occur simultaneously, in contrast to the 3D case \cite{cite-27, cite-28, cite-29, cite-31,cite-32}, or pure 2D superfluid state \cite{cite-30}. Furthermore, we show that when increasing the interlayer Coulomb interaction parameter, the excitonic condensate state is stabilizing, and the DOS spectra become gapless for all temperature ranges. Therefore, we suggest that the system is always in the mixed state in this case, and the excitonic condensation and pair formation are coexisting. We give also the temperature dependence of all calculated quantities presented in the work. As the unit of energy, we have chosen the intralayer hopping amplitude for the layer-1 and we put $|t_{1}|=1$. We set also $k_{B}=1$ and $\hbar=1$ overall in the paper and for the 2D square lattice constant $a$ we have putted $a\equiv 1$. 

The paper is organized as follows: in the Section \ref{sec:Section_2} we introduce the Hamiltonian of the model with the effective chemical potential. In the Section \ref{sec:Section_3} we derive the expression of the general action in the system and the general form of the partition function. In the Section \ref{sec:Section_4} we present the solution for the chemical potential in the DL system. The Section \ref{sec:Section_5} is entirely devoted to the single-particle excitonic spectral functions and DOS evaluations. Both, analytical and numerical results are given there. In the Section \ref{sec:Section_6} we discuss our results in touch with the experimental accessibilities, and in the Section \ref{sec:Section_7} we give a conclusion. Finally, in the Appendix \ref{Section_8}, we present the calculation details for the DOS functions. 
%
\section{\label{sec:Section_2} The model}
%
We introduce the Hubbard Hamiltonian of our DL system, with the intraplane and interplane Coulomb couplings. The DL is composed of two square lattices, numbered as $l=1,2$ and doped respectively with electrons and holes. The Hamiltonian reads as 
\begin{eqnarray}
H=-\sum_{\substack{\left\langle i, j \right\rangle\\ l,\sigma}}t_{l}\left(c^{\dag}_{li,\sigma}c_{lj,\sigma}+h.c.\right)-t_{\perp}\sum_{i,\sigma}\left(c^{\dag}_{1i,\sigma}c_{2i,\sigma}+h.c.\right)
\nonumber\\
-\sum_{i,l,\sigma}\bar{\mu}_{l}n_{li,\sigma}+U\sum_{i,l}\left(\frac{n^{2}_{li}}{4}-S^{2}_{lz}\right)+W\sum_{i,\sigma,\sigma'}n_{1i,\sigma}n_{2i,\sigma'}.
\label{Equation_1}
\end{eqnarray}
Here, $\left\langle i, j \right\rangle$ denotes the sum over the nearest neighbors lattice sites, $t_{l}$ $l=1, 2$ is the intraplane hopping amplitude, $\sigma$ denotes the spins of the electrons with two possible polarization directions ($\sigma= \uparrow, \downarrow$), and $t_{\perp}$ is the interplane hopping amplitude. Furthermore, $\bar{\mu}_{l}$ $l=1,2$ are the shifted chemical potentials
\begin{eqnarray}
\bar{\mu}_{l}=\mu_{l}+U/2+W,
\label{Equation_2}
\end{eqnarray}
and $W$ is the local interlayer Coulomb repulsion. Next, $n_{li}=n_{li,\uparrow}+n_{li,\downarrow}$ is the total electron density on site $i$, in the layer $l$, and $S_{lz}$ is the $z$-component of the generalized spin operator
\begin{eqnarray}
S_{lz}=\frac{1}{2}\left(n_{li,\uparrow}-n_{li,\downarrow}\right).
\label{Equation_3}
\end{eqnarray}
For a simple treatment, we suppose that the chemical potentials in both layers are opposite in sign $\mu_{1}=-\mu_{2}$. This, in turn, will introduce the p-doping in one layer and n-doping into another. The parameter $U$ in Eq.(\ref{Equation_1}) is the local intralayer Coulomb interaction. We consider the double layer with respect to the half-filling conditions in each layer $\left\langle n_{l} \right\rangle=1$, for $l=1,2$. The formation of the excitons in the layered structure, and the possibility of their further condensation at the low temperatures requires the attraction between the electrons and holes. This is described by the Hubbard interaction term between different layers and is given by the last term in Eq.(\ref{Equation_1}).

Next, we will pass to the Grassmann representation for the fermionic variables, and we write the partition function of the system, by employing the imaginary time fermion path integral method \cite{cite-34}.
For this, we introduce the imaginary-time variables $\tau$, at each lattice site $i$. The time variables $\tau$ vary in the interval $(0,\beta)$, where $\beta=1/T$ with $T$ being the temperature. The time-dependent Grassmann variables $c_{li,\sigma}(\tau)$ (${c}^{\dag}_{li,\sigma}(\tau)$) are satisfying the anti-periodic boundary conditions for fermions \cite{cite-35}. Then, the grand canonical partition function of the system is 
\begin{eqnarray}
Z=\int\left[Dc^{\dag}Dc\right]e^{-S\left[c^{\dag},c\right]},
\label{Equation_4}
\end{eqnarray}
where, the action in the exponent is expressed as
\begin{eqnarray}
S\left[c^{\dag},c\right]=S_{\rm B}\left[c^{\dag},c\right]+\int^{\beta}_{0}d\tau H\left(\tau\right).
\label{Equation_5}
\end{eqnarray}
The first term in Eq.(\ref{Equation_5}), is the fermionic Berry-term. It is given as
\begin{eqnarray}
S_{\rm B}\left[c^{\dag},c\right]=\sum_{i,l,\sigma}\int^{\beta}_{0}d\tau c^{\dag}_{li,\sigma}(\tau)\frac{\partial}{\partial \tau}c_{li,\sigma}(\tau).
\label{Equation_6}
\end{eqnarray}
Furthermore, we will combine the quadratic and linear total density terms in Eq.(\ref{Equation_1}) and we will decouple the obtained nonlinear term by the scalar field Hubbard-Stratanovich linearization procedure \cite{cite-34}. 

\section{\label{sec:Section_3} The method}

For further simplifications, we will write the interlayer interaction
term in Eq.(\ref{Equation_1}) in a compact form, by introducing
the following complex variables $\xi_{i,\sigma\sigma'}(\tau)=c^{\dag}_{2i,\sigma}(\tau)c_{1i,\sigma'}(\tau)
$, and their complex conjugates, with all possible spin directions. The variables $\xi_{i,\sigma\sigma'}(\tau)$ are linear in the density of electrons. Then, the interlayer Coulomb interaction term reads as 
\begin{eqnarray} S\left[\xi^{\dag},\xi\right]=-W\sum_{i,\sigma,\sigma'}\int^{\beta}_{0}d\tau|\xi_{i,\sigma\sigma'}(\tau)|^{2}.
\label{Equation_7}
\end{eqnarray}
The procedures of decoupling the nonlinear interaction terms are rather standard \cite{cite-28, cite-29, cite-34} and we will not present here the details. We will just write the form of the partition function after the decoupling procedure
\begin{eqnarray}
Z=\int\left[Dc^{\dag}Dc\right]\int\left[D\Delta^{\dag}D\Delta\right]e^{-S\left[c^{\dag},c,\Delta^{\dag}, \Delta\right]},
\label{Equation_8}
\end{eqnarray}
where $\Delta^{\dag}$ and $\Delta$ are the new decoupling fields related to the complex variables $\xi^{\dag}_{i,\sigma\sigma'}(\tau)$ and $\xi_{i,\sigma\sigma'}(\tau)$, and the total action of the system contains already the scalar saddle-point terms appearing after the decoupling 
\begin{eqnarray}
&&S\left[c^{\dag},c,\Delta^{\dag}, \Delta\right]=S_{\rm B}[c^{\dag},c]+\sum_{\substack{\left\langle i, j \right\rangle\\ l,\sigma}}\int^{\beta}_{0}d\tau t_{l}\left(c^{\dag}_{li,\sigma}(\tau)c_{lj,\sigma}(\tau)+h.c.\right)
\nonumber\\
&&+t_{\perp}\sum_{i,\sigma}\int^{\beta}_{0}d\tau\left(c^{\dag}_{1i,\sigma}(\tau)c_{2i,\sigma}(\tau)+h.c.\right)
+S\left[V^{S}\right]+S\left[\Delta^{S}_{c}\right]+S\left[\xi^{\dag},\xi,\Delta^{\dag},\Delta\right].
\label{Equation_9}
\end{eqnarray}
Here, 
\begin{eqnarray}
S\left[\Delta^{S}_{c}\right]=-\sum_{i,l,\sigma}\int^{\beta}_{0}d\tau (-1)^{\sigma}\Delta^{S}_{lc}n_{li,\sigma},
\label{Equation_10}
\end{eqnarray}
where $\Delta^{S}_{lc}$ is the saddle-point value of the scalar parameter related to the decoupling of the electron density-difference quadratic term in the Hamiltonian in Eq.(\ref{Equation_1}). Considering the case of half-filling, we will fix $\Delta^{S}_{1c}=\Delta^{S}_{2c}=U/2$. The action $S\left[\xi^{\dag},\xi,\Delta^{\dag},\Delta\right]$ in Eq.(\ref{Equation_9}) reads then as
\begin{eqnarray}
&&S\left[\xi^{\dag},\xi,\Delta^{\dag},\Delta\right]=\sum_{i,\sigma,\sigma'}\int^{\beta}_{0}d\tau\left[\frac{|\Delta_{i,\sigma\sigma'}(\tau)|^{2}}{W}+\Delta^{\dag}_{i,\sigma\sigma'}(\tau)\xi_{i,\sigma\sigma'}(\tau)+\xi^{\dag}_{i,\sigma\sigma'}(\tau)\Delta_{i,\sigma\sigma'}(\tau)\right].
\label{Equation_11}
\end{eqnarray}
Furthermore, we see that the decoupling fields $\Delta^{\dag}_{i,\sigma\sigma'}(\tau)$ and $\Delta_{i,\sigma\sigma'}(\tau)$ are descendants of the pairing gap parameters for different spin orientations. Indeed, the saddle point expressions for
the decoupling fields are 
\begin{eqnarray}
\Delta^{S}_{\sigma\sigma'}=-W\left\langle c^{\dag}_{2i,\sigma}\left(\tau\right)c_{1i,\sigma'}\left(\tau\right)\right\rangle,
\label{Equation_12}
\newline\\
\Delta^{\dag S}_{\sigma\sigma'}=-W\left\langle c^{\dag}_{1i,\sigma}\left(\tau\right)c_{2i,\sigma'}\left(\tau\right)\right\rangle.
\label{Equation_13}
\end{eqnarray}   

The quantum statistical averages here are given with the help of the fermionic action in Eq.(\ref{Equation_9}), mainly
\begin{eqnarray}
\left\langle ... \right\rangle=Z^{-1}\int\left[Dc^{\dag}Dc\right]\int\left[D\Delta^{\dag}D\Delta\right]...e^{-S\left[c^{\dag},c,\Delta^{\dag}, \Delta\right]}.
\label{Equation_14}
\end{eqnarray}
In fact, the parameter $\Delta^{S}_{\sigma\sigma'}$ corresponds to the conventional excitonic pairing gap, as in the usual bulk semiconductors \cite{cite-27,cite-28, cite-29, cite-36,cite-37,cite-38,cite-39,cite-40}.

Without any loss of generality, we suppose the case of the pairing states with the uniform real gap parameters \cite{cite-28, cite-29, cite-30, cite-40}. It is not difficult to write the total action of the system, given in Eq.(\ref{Equation_9}), in the Fourier transformed form
\begin{eqnarray}
S\left[\Psi^{\dag},\Psi,\Delta^{\dag},\Delta\right]=\frac{1}{\beta{N}}\sum_{{\bf{k}},\nu_{n}}\Psi^{\dag}_{{\bf{k}}}(\nu_{n})\hat{G}^{-1}_{{\bf{k}}}\left(\nu_{n}\right)\Psi_{{\bf{k}}}(\nu_{n}),
\label{Equation_15}
\end{eqnarray}
where $N$ is the total number of $k$-points in the first Brillouin zone (BZ), and we have introduced the 4-component Nambu spinor 
\begin{eqnarray}
\Psi_{{\bf{k}}}(\nu_{n})=\left[
c_{1{\bf{k}},\uparrow}(\nu_{n}), c^{\dag}_{1{\bf{k}},\downarrow}(\nu_{n}), c_{2{\bf{k}},\downarrow}(\nu_{n}), c^{\dag}_{2{\bf{k}},\uparrow}(\nu_{n})
\right]^{T}
\label{Equation_16}
\end{eqnarray}
and the complex conjugate Nambu vector $\Psi^{\dag}_{{\bf{k}}}(\nu_{n})$. Here ${\bf{k}}$, is the electron wave vector in the reciprocal Fourier space, and the indexes, before the wave vectors, in Eq.(\ref{Equation_16}) indicate the layer they are connected to. The fermionic Matsubara ``odd'' frequencies \cite{cite-35} are $\nu_{n}={(2n+1)\pi/\beta}$, with $n=0,\pm 1,\pm2,...$ .   

The matrix $\hat{G}^{-1}_{{\bf{k}}}\left(\nu_{n}\right)$, in Eq.(\ref{Equation_15}), is a 4 $\times$ 4 matrix, and is given as 
\begin{eqnarray}
\hat{G}^{-1}_{{\bf{k}}}\left(\nu_{n}\right)=\left(
\begin{array}{ccccrrrr}
E_{1,{\bf{k}}}\left(\nu_n\right) & 0 & {t}_{\perp}+\Delta_{\uparrow\uparrow} & 0\\
0 &-E_{1,{\bf{k}}}\left(\nu_n\right)  & 0 & -{t}_{\perp}-\Delta^{\dag}_{\downarrow\downarrow} \\
{t}_{\perp}+\Delta^{\dag}_{\uparrow\uparrow} & 0 & E_{2,{\bf{k}}}\left(\nu_n\right) & 0 \\
0 & -{t}_{\perp}-\Delta_{\downarrow\downarrow} & 0 & -E_{2,{\bf{k}}}\left(\nu_n\right)  
\end{array}
\right).
\label{Equation_17}
\end{eqnarray}
The energy parameters $E_{l,{\bf{k}}}\left(\nu_n\right)$ $l=1,2$, entering into the Eq.(\ref{Equation_17}), are given by
\begin{eqnarray}
E_{l,{\bf{k}}}\left(\nu_n\right)=-i\nu_{n}-\Delta^{S}_{lc}+2\tilde{t}_{l,{\bf{k}}}+\bar{\mu}_{l}-\frac{U\bar{n}_{l}}{2}.
\label{Equation_18}
\end{eqnarray}
Next, $\tilde{t}_{l,{\bf{k}}}$ $l$=1,2, are the renormalized, intraplane hopping amplitudes 
\begin{eqnarray}
\tilde{t}_{l,{\bf{k}}}=2t_{l}\gamma_{\bf{k}},
\label{Equation_19}
\end{eqnarray}
and the energy dispersion $\gamma_{\bf{k}}$, in Eq.(\ref{Equation_19}), is
\begin{eqnarray}
\gamma_{\bf{k}}=\cos(k_{x}a)+\cos(k_{y}a),
\label{Equation_20}
\end{eqnarray}
where $a$ is the 2D lattice constant, and we will put $a\equiv 1$ as it is mentioned in the introduction of the present paper. 
The total density average $\bar{n}_{l}$, for the layer $l$, is given with the help of the action in Eq.(\ref{Equation_9}) or in Eq.(\ref{Equation_15}).

Then, for the partition function of the system, we have
\begin{eqnarray} Z=\int\left[D\Psi^{\dag}D\Psi\right]\int\left[D\Delta^{\dag}D\Delta\right]e^{-S\left[\Psi^{\dag},\Psi,\Delta^{\dag},\Delta\right]}.
\label{Equation_21}
\end{eqnarray}
This form of the partition function will be used for calculating the Green functions, normal spectral functions, and DOS spectra.

\section{\label{sec:Section_4} The behavior of the chemical potential}
\begin{figure}
\begin{center}
\includegraphics[width=250px,height=250px]{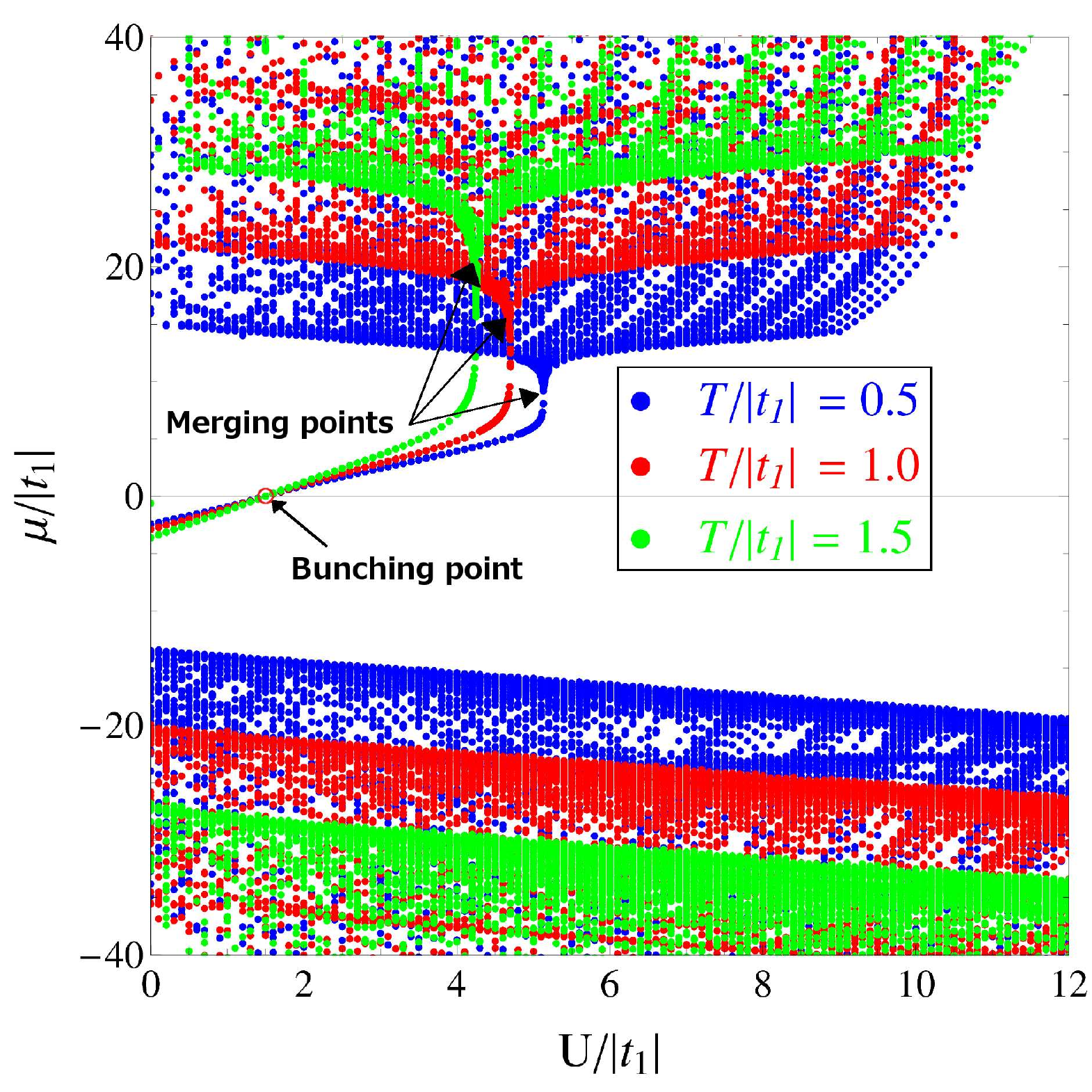}
\caption{\label{fig:Fig_1}(Color online) 
The temperature dependence of the chemical potential in the DL structure is shown. The interlayer Coulomb interaction parameter is fixed at $W/|t_{1}|=0.75$ and $t_{\perp}=0.04|t_{1}|$. The beginning of the Frenkel channel solution (see in Ref.\onlinecite{cite-33}) is displaced to the smaller intralayer Coulomb interaction region.}
\end{center}
\end{figure}

%
From the structure of the inverse Green function matrix, given in Eq.(\ref{Equation_17}), it follows that $\Delta_{\sigma\sigma'}=\Delta_{\sigma\sigma}\delta_{\sigma\sigma'}$. We should mention here that this result is true only for the DL system without an applied external electric field, such as in our consideration. For the general case, when an external electric field is present, then $\Delta^{\rm ex}_{\uparrow\uparrow}\neq\Delta^{\rm ex}_{\downarrow\downarrow}$, because of the imbalance $|\mu_{1}|\neq|\mu_{2}|$. We will not meditate here this case, and it will be the subject of our future considerations. Taking into account the half-filling condition for the  total average particle density in each layer, we have found the following system of coupled self-consistent nonlinear equations   
\begin{eqnarray}
&&\frac{1}{N}\sum_{{\bf{k}}}\left[n_{F}\left(\kappa_{1,{\bf{k}}}\right)+n_{F}\left(\kappa_{2,{\bf{k}}}\right)\right]=1,
\label{Equation_22}
\newline\\
&&\Delta=-\frac{W}{2N}\sum_{{\bf{k}}}\frac{\left({t}_{\perp}+\Delta\right)\cdot\left[n_{F}\left(\kappa_{1,{\bf{k}}}\right)+n_{F}\left(\kappa_{2,{\bf{k}}}\right)\right]}{\sqrt{\eta^{2}_{{\bf{k}}}+\left({t}_{\perp}+\Delta\right)^{2}}},
\label{Equation_23}
\end{eqnarray}
where $n_{F}\left(z\right)$ is the Fermi-Dirac distribution function $n_{F}\left(z\right)=1/(e^{\beta{z}}+1)$, and the parameter $\eta_{{\bf{k}}}$ is defined 
as $\eta_{{\bf{k}}}=\tilde{t}_{1{\bf{k}}}-\tilde{t}_{2{\bf{k}}}-\mu$.
For the convenience, we have omitted the spin indexes near the excitonic gap parameter $\Delta$.
Furthermore, the parameters $\kappa_{i,{\bf{k}}}$ $i=1,2$ in Eqs.(\ref{Equation_22}) and (\ref{Equation_23}), entering in the argument of the distribution function, are given explicitly by the following relations
\begin{eqnarray}
\kappa_{i,{\bf{k}}}=-\frac{U}{2}+2\left(t_{1}+t_{2}\right)\gamma_{\bf{k}}+W+(-1)^{i-1}\sqrt{\eta^{2}_{{\bf{k}}}+\left({t}_{\perp}+\Delta\right)^{2}}.
\label{Equation_24}
\end{eqnarray}
We have solved the system of equations in Eqs.(\ref{Equation_22}) and (\ref{Equation_23}) for the case of the electron hopping-asymmetry (i.e., $t_{1}\neq t_{2}$ and $t_{1}<0$ for hole layer) in different layers of the system. We will not present here the results for the excitonic gap parameter $\Delta$, and we refer the reader to the work in Ref.\onlinecite{cite-33} for a detailed discussion. We concentrate here on the chemical potential solution. Particularly, it is shown in Ref.\onlinecite{cite-33} that the interlayer Coulomb interaction parameter $W/|t_{1}|$ has a more important impact on the amplitude of the excitonic gap than the interlayer hopping parameter $t_{\perp}$, which leads to the stabilization of the quantum coherent states, such in the case of the valence bond solids \cite{cite-41} and supersolids \cite{cite-42, cite-43, cite-44} resulting from the strong interplane attractions (in the case of the heavy fermions, the role of the interlayer hopping amplitude is negligibly small). 
Namely, from the results, given in  Ref.\onlinecite{cite-33}, it follows that both gap channels: Frenkel type, with strongly bound excitons, and Wannier-Mott (WM) type, with weakly bound excitons (see the lowest pairing gap solutions in Figs.~\ref{fig:Fig_1} and ~\ref{fig:Fig_2} in Ref.\onlinecite{cite-33}) are found to be possible in the correlated DL system. For strong intralayer Coulomb interaction values, the Frenkel-gap disappears, and the fluctuating WM gap remains open until very high values of the interaction parameter $U/|t_{1}|$ \cite{cite-33}.
%
%
%
In Fig.~\ref{fig:Fig_1}, we have presented the solution for the chemical potential in the case of weak interlayer Coulomb interaction parameter $W/|t_{1}|=0.75$, and $t_{\perp}=0.04|t_{1}|$. The accuracy of convergence for numerical solutions is achieved with a relative error of the order $10^{-7}$, and with an absolute error of the order $10^{-4}$. We see in Fig.~\ref{fig:Fig_1} a spectacular behavior of the chemical potential, where the merging point corresponds to the merging of two type excitonic solutions. The wire solutions of $\mu$ corresponds to the Frenkel type excitonic channel, while the rest is due to the WM gap solution (see the low-energy gap solutions in Ref.\onlinecite{cite-33}). In addition to the results, given in Ref.\onlinecite{cite-33}, here, it is important to mention the existence of a stable ``Bunching-point'' Frenkel solution of the chemical potential (see Fig.~\ref{fig:Fig_1}), where different solutions of $\mu$, corresponding to Frenkel channels and at different temperatures, are intersecting, and the position of this point on the $U/|t_{1}|$-axis is still unchanged, when varying the temperature. This point corresponds to the intraplane Coulomb interaction value $U_{B}/|t_{1}|=1.5$, for the case discussed in Fig.~\ref{fig:Fig_1}. The role of this stable solution will be clear hereafter when discussing the normal spectral functions, and DOS properties at $U_{B}/|t_{1}|=1.5$. 
%
\section{\label{sec:Section_5} Spectral functions and single-particle DOS}
%
\renewcommand\thesubsection{\thesection.\arabic{subsection}}
%
\subsection{\label{sec:Section_5_1} Normal and anomalous spectral functions}
%
For the layer $l$, we define the equal time, normal electron Matsubara Green functions \cite{cite-32} as:
\begin{eqnarray}
G^{\sigma,\sigma}_{l}(i\tau,i\tau)=-\left\langle c^{\dag}_{li,\sigma}(\tau)c_{li,\sigma}(\tau)\right\rangle=\frac{1}{\beta{N}}\sum_{{\bf{k}},\nu_{n}}G^{\sigma,\sigma}_{l,{\bf{k}}}(\nu_{n}),
\label{Equation_25}
\end{eqnarray}
where
\begin{eqnarray}
G^{\sigma,\sigma}_{l,{\bf{k}}}(\nu_{n})=-\frac{1}{\beta{N}}\left\langle c^{\dag}_{l{\bf{k}},\sigma}(\nu_{n})c_{l{\bf{k}},\sigma}(\nu_{n})\right\rangle.
\label{Equation_26}
\end{eqnarray}
Here, the average should be done with the help of the action given in Eq.(\ref{Equation_15}).
Then, we will fix the spin $\sigma=\uparrow$ and we omit the spin indexes near the Green functions. The case $\sigma=\downarrow$ is identical and gives the same results, because of the symmetry of the considered DL system. Using the expression of the inverse Green function matrix, given in Eq.(\ref{Equation_17}), we will have for the functions $G_{l,{\bf{k}}}(\nu_{n})$ of the layers $l=1,2$
\begin{eqnarray}
G_{l,{\bf{k}}}(\nu_{n})=-\frac{-i\nu_{n}-\Delta^{S}_{c}+4t_{\tilde{l}}\gamma_{{\bf{k}}}+\bar{\mu}_{\tilde{l}}-\frac{U\bar{n}_{\tilde{l}}}{2}}{(i\nu_{n})^{2}+2i\nu_{n}a_{{\bf{k}}}+b_{{\bf{k}}}-(t_{\perp}+\Delta)^{2}},
\label{Equation_27}
\end{eqnarray}
where $\tilde{l}$ means the layer opposite to the layer $l$, and the ${\bf{k}}$-dependent parameters $a_{{\bf{k}}}$ and $b_{{\bf{k}}}$ in Eq.(\ref{Equation_27}) are defined as follows
\begin{eqnarray}
a_{{\bf{k}}}=\Delta^{S}_{c}-2(t_{1}+t_{2})\gamma_{{\bf{k}}}-W,
\label{Equation_28}
\newline\\
b_{{\bf{k}}}=(\Delta^{S}_{c}-4t_{1}\gamma_{{\bf{k}}}-\tilde{\mu}_{1})(\Delta^{S}_{c}-4t_{2}\gamma_{{\bf{k}}}-\tilde{\mu}_{2}).
\label{Equation_29}
\end{eqnarray}
The effective chemical potentials $\tilde{\mu}_{l}$ $l=1,2$, introduced here, are given by
\begin{eqnarray}
\tilde{\mu}_{l}=\bar{\mu}_{l}-\frac{U\bar{n}_{l}}{2}.
\label{Equation_30}
\end{eqnarray}
As it is discussed above, we will put $\Delta^{S}_{lc}=\frac{1}{2}$, and $\bar{n}_{l}=1$ for the layers $l=1,2$. 

In the simple Angle Resolved Photoemission Spectroscopy (ARPES) experiments, the direct detection quantities are the single-particle spectral functions $A_{l,{\bf{k}}}(\nu)$, which gives the information about the occupation
of the state ${\bf{k}}$ at the given frequency $\nu$. Thus, we will evaluate these functions for our model. The single-particle spectral functions $A_{l,{\bf{k}}}(\nu)$ $l=1,2$ are defined with the help of the imaginary part of the retarded Matsubara Green functions \cite{cite-35}
\begin{eqnarray}
A_{l,{\bf{k}}}(\nu)=-\frac{1}{\pi}\Im G_{l,{\bf{k}}}(\nu_{n})|_{i\nu_{n}\rightarrow \nu+i\alpha^{+}},
\label{Equation_31}
\end{eqnarray}
where the analytical continuation is performed into the upper-half complex semi-plane, and $\alpha^{+}$ is the infinitesimal positive constant.
Using the expression of the $l$-layer normal Green function, given in Eq.(\ref{Equation_27}), it is not difficult to show that the $l$-layer single-particle normal spectral function $A_{l,{\bf{k}}}(\nu_{n})$ takes the following form
\begin{eqnarray}
A_{l,{\bf{k}}}(\nu)=&&\sum_{i=1,2}\frac{(-1)^{l+i}}{2}\left[\frac{\eta_{{\bf{k}}}}{\sqrt{D_{\bf{k}}}}-(-1)^{l+i-1}\frac{}{}\right]\delta_{\rm L}(\alpha,\nu,\kappa_{i,{\bf{k}}}).
\label{Equation_32}
\end{eqnarray}
Here, the functions $\eta_{{\bf{k}}}$ and $D_{\bf{k}}$ are defined in the following way
\begin{eqnarray}
\eta_{{\bf{k}}}=2(t_{1}-t_{2})\gamma_{\bf{k}}-\mu,
\label{Equation_33}
\newline\\
D_{\bf{k}}=\eta^{2}_{\bf{k}}+(t_\perp+\Delta)^{2}.
\label{Equation_34}
\end{eqnarray}
The function $\delta_{\rm L}(\alpha,\nu,\kappa_{i,{\bf{k}}})$ in Eq.(\ref{Equation_32}) is the Lorentzian broadening function with the sufficiently small positive broadening parameter $\alpha$. It is given as 
$\delta_{\rm L}(\alpha,\nu,\kappa_{i,{\bf{k}}})=\frac{1}{\pi}{\alpha}/\left[{\alpha^{2}+(\nu-\kappa_{i,{\bf{k}}})^{2}}\right],$
%
\begin{figure}
\begin{center}
\includegraphics[width=260px,height=450px]{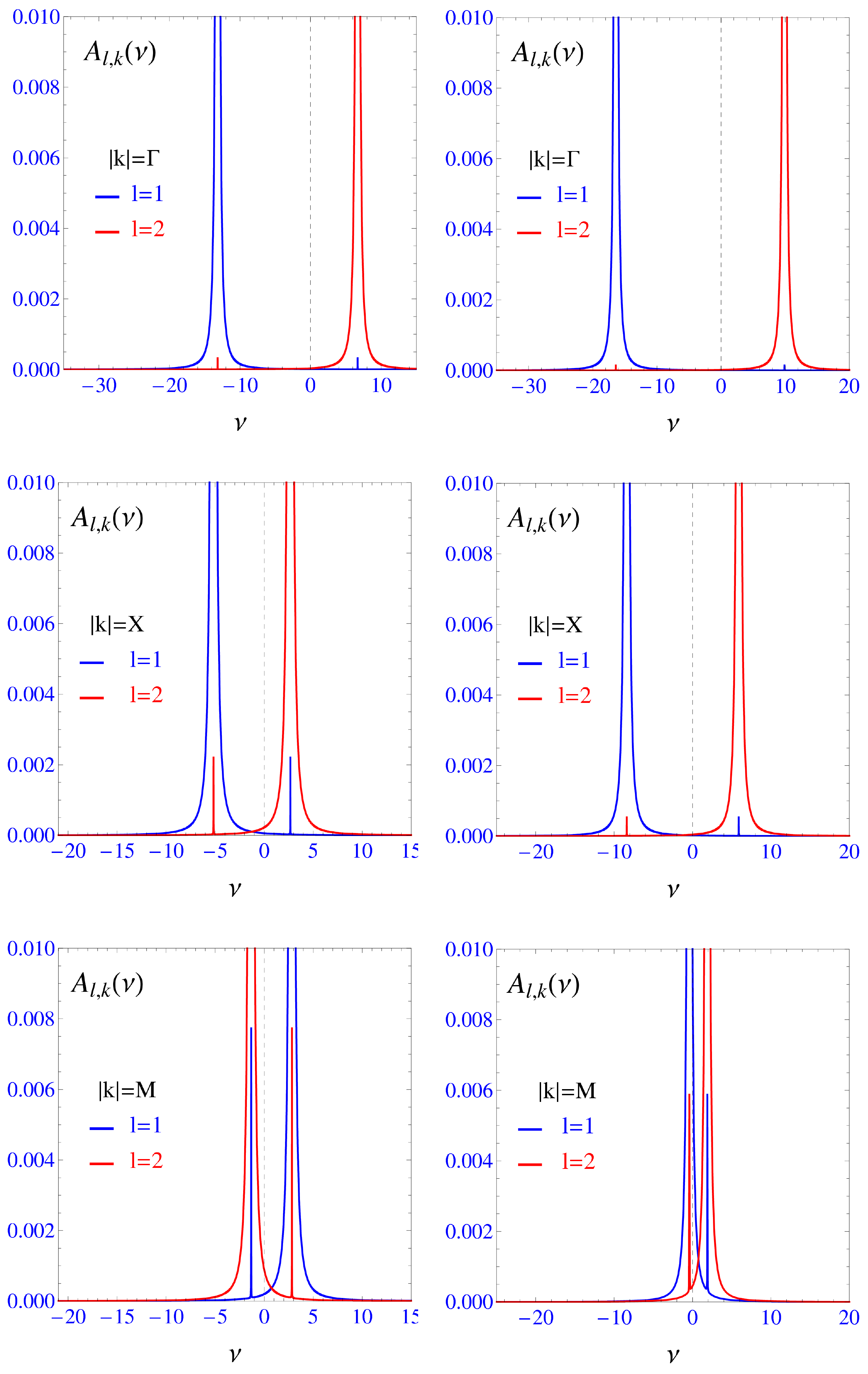}
\caption{\label{fig:Fig_2}(Color online) The excitonic normal spectral function $A_{l,{\bf{k}}}(\nu)$, along the high-symmetry direction $\Gamma\rightarrow M$. The hole (see the curves in blue) and electron (see the curves in red) layers $l=1,2$ are considered, and the temperature is fixed at $T/|t_{1}|=0.5$ (for the left panel) and at $T/|t_{1}|=1.5$ (for the right panel).}
\end{center}
\end{figure} 
%
\begin{figure}
\begin{center}
\includegraphics[scale=.5]{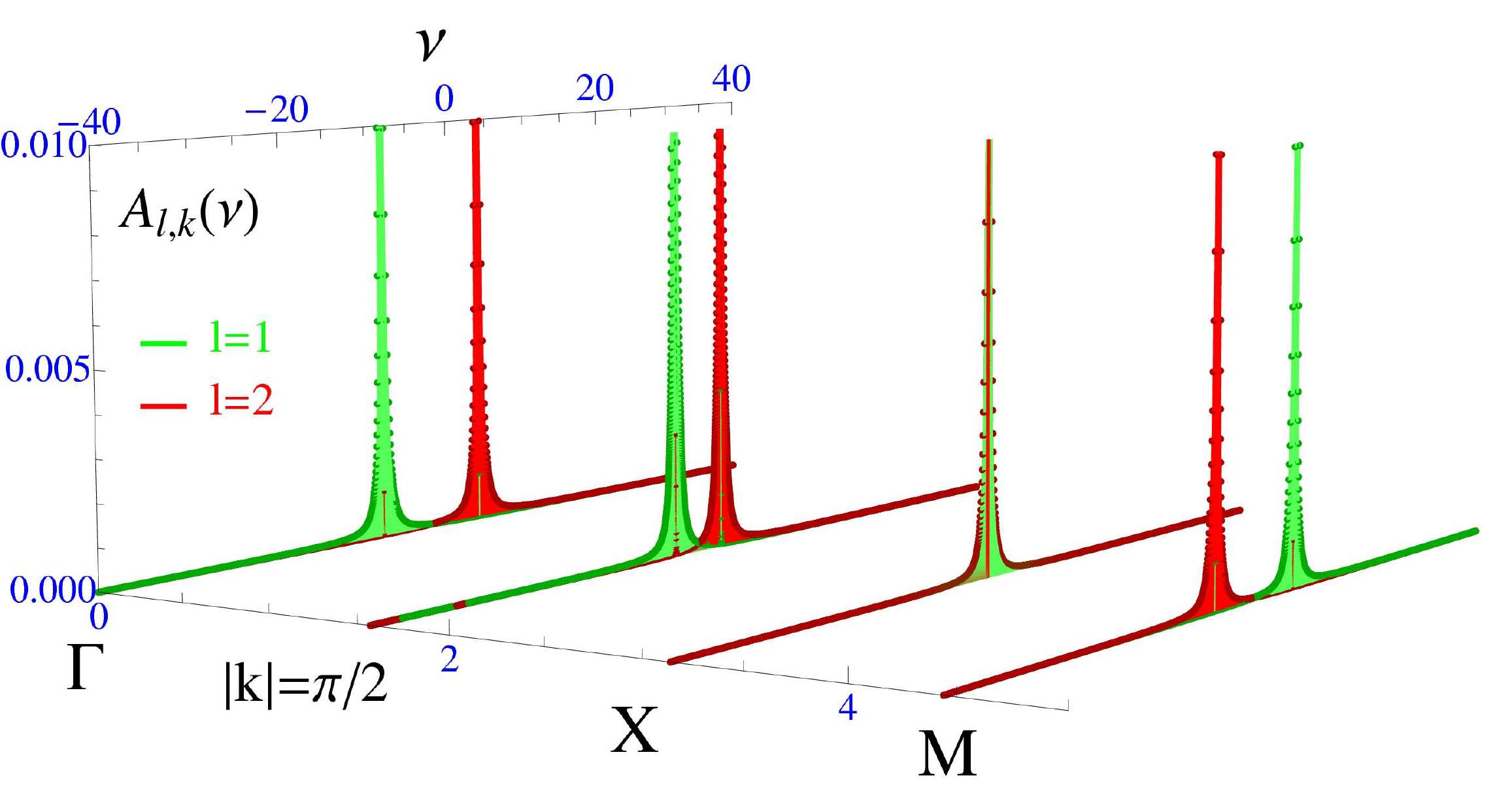}
\caption{\label{fig:Fig_3}(Color online) The excitonic normal spectral function $A_{l,{\bf{k}}}(\nu)$ at the high-symmetry points $\Gamma$, $X$ and $M$ for the bunching point of the chemical potential (see the $\mu$-structure in Fig.~\ref{fig:Fig_1}). The hole (see the curves in green) and electron (see the curves in red) layers $l=1,2$ are considered, and the temperature is fixed at $T/|t_{1}|=0.5$. The spectral function collapse is clearly seen at the symmetry point $X$.}
\end{center}
\end{figure} 
%
where the parameters $\kappa_{i,{\bf{k}}}$ are given in Eq.(\ref{Equation_24}).
The numerical evaluations of the spectral functions $A_{l,{\bf{k}}}(\nu)$, for the layers $l=1,2$, are presented in Figs.~\ref{fig:Fig_2} and ~\ref{fig:Fig_3}. The broadening parameter $\alpha$ is chosen very small $\alpha=0.005$. In Fig.~\ref{fig:Fig_2}, the normal spectral functions for both layers, $l=1,2$, are presented at the high-symmetry points $\Gamma$, $X$, and $M$, on the 2D square lattice. The interlayer Coulomb interaction parameter is set at $W/|t_{1}|=0.75$ and $t_{\perp}=0.04|t_{1}|$. The intralayer Coulomb interaction $U/|t_{1}|$ is chosen away from the value at the Bunching-point: $U_{\rm B}/|t_{1}|=1.5$ (see in Fig.~\ref{fig:Fig_1}) for this case. The left panel in Fig.~\ref{fig:Fig_2} corresponds to $T/|t_{1}|=0.5$, and the curves in the right panel are evaluated for the higher temperature $T/|t_{1}|=1.5$. We see in the left panel in Fig.~\ref{fig:Fig_2} that the normal spectral functions lines for the layers $l=1,2$ are interchanging with their positions on the frequency axis at the symmetry point ${\bf{k}}=M$. For the higher temperature case (see the curves in the right panel in Fig.~\ref{fig:Fig_2}), the separation between the spectral lines in different layers is considerably increased, and the mentioned interchanging does not takes place in this case. 

In Fig.~\ref{fig:Fig_3}, the Frenkel spectral functions evaluations for the layers $l=1,2$ are given at the Bunching point $U_{\rm B}/|t_{1}|=1.5$ (see in Fig.~\ref{fig:Fig_1}) and projected on the plane $(\nu,|{\bf{k}}|)$. The temperature is set at $T/|t_{1}|=0.5$. We see in Fig.~\ref{fig:Fig_3} that at the symmetry point $|{\bf{k}}|=X$, an unusual collapse occurs in the spectral functions structure, which is furthermore removed when evaluating along the symmetry direction $X\rightarrow M$.     
%
\begin{figure}
\begin{center}
\includegraphics[scale=.3]{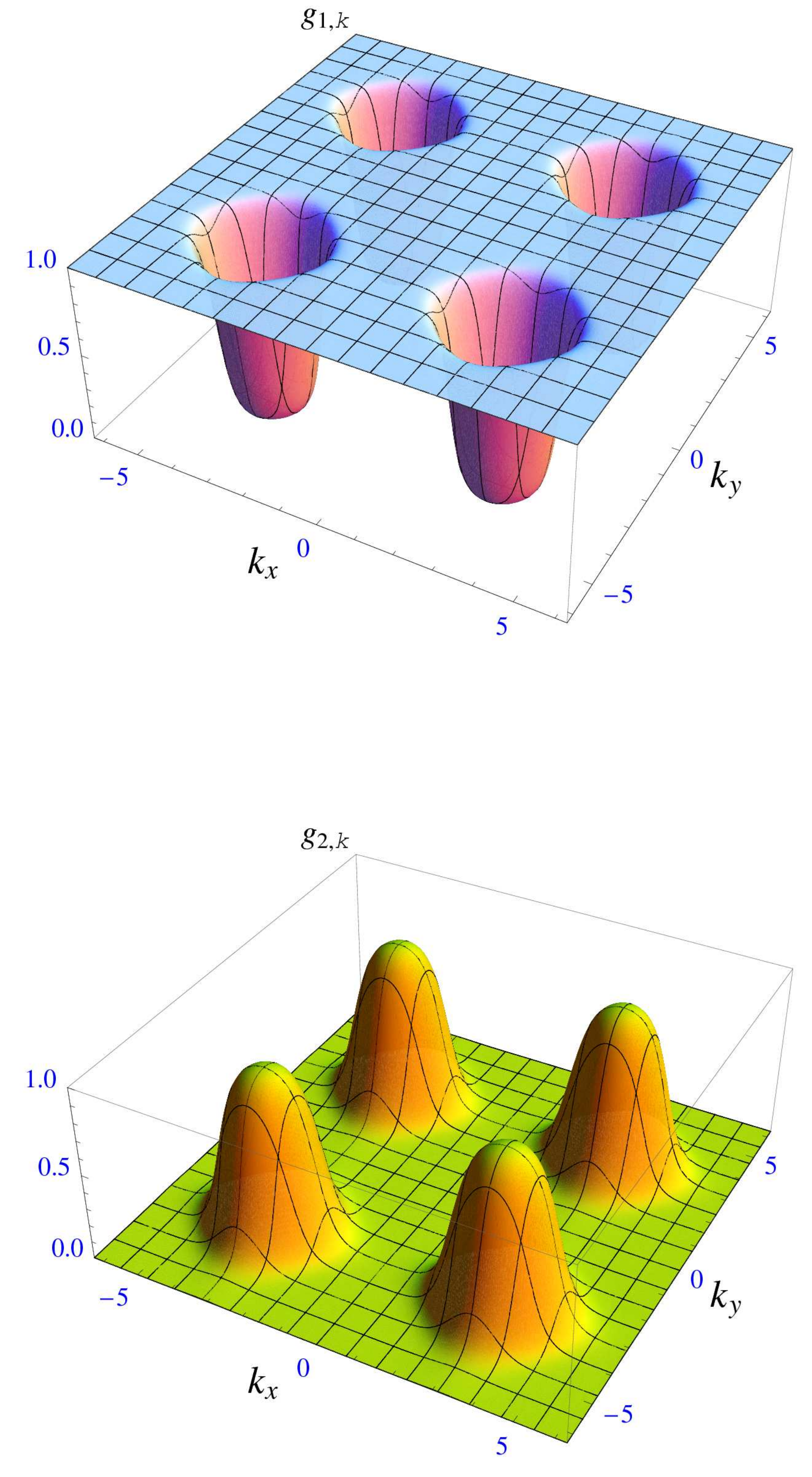}
\caption{\label{fig:Fig_4}(Color online) The plot of the frequency-summed function $g_{l,{\bf{k}}}$ as a function of the ${\bf{k}}=k_{x}{\bf{G}}_{1}+k_{y}{\bf{G}}_{2}$ wave vector, in the extended BZ scheme. The upper panel shows the ${\bf{k}}$ dependence for the hole layer, with $l=1$, and, in the lower panel, the ${\bf{k}}$ dependence for the electron layer, with $l=2$ is shown. The temperature is set at $T/|t_{1}|=0.5$.}
\end{center}
\end{figure} 
%
\begin{figure}
\begin{center}
\includegraphics[scale=.3]{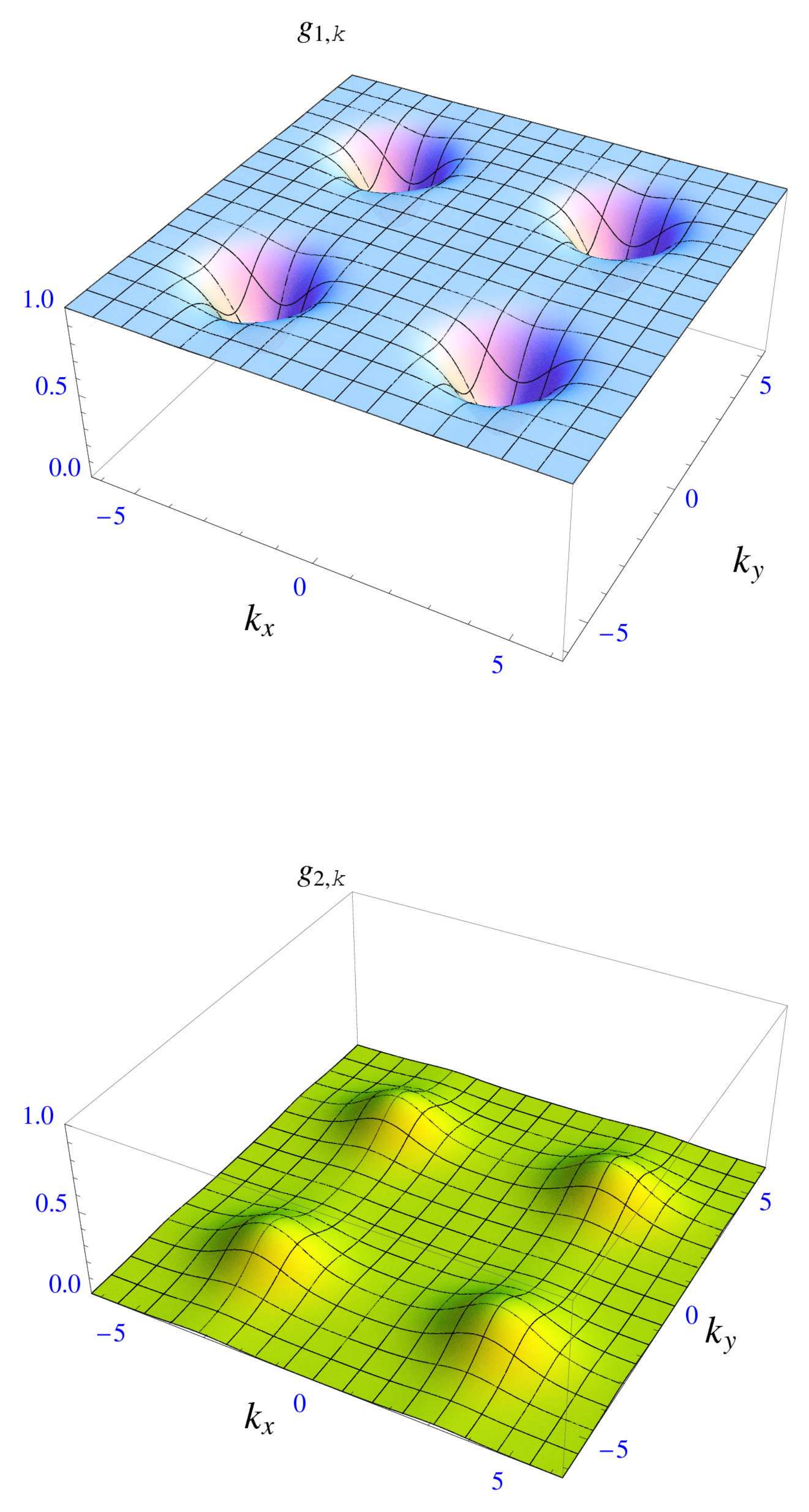}
\caption{\label{fig:Fig_5}(Color online)The plot of the frequency-summed function $g_{l,{\bf{k}}}$ as a function of the ${\bf{k}}=k_{x}{\bf{G}}_{1}+k_{y}{\bf{G}}_{2}$ wave vector, in the extended BZ scheme. The upper panel shows the ${\bf{k}}$ dependence for the hole layer, with $l=1$, and, in the lower panel, the ${\bf{k}}$ dependence for the electron layer, with $l=2$ is shown. The temperature is set at $T/|t_{1}|=1.5$.}
\end{center}
\end{figure} 
%
Next, we have evaluated the frequency-summed normal Green functions $g_{l,{\bf{k}}}$, for the layers $l=1,2$. For this, we have performed the fermionic Matsubara summation in Eq.(\ref{Equation_27}). We get
\begin{eqnarray}
g_{l,{\bf{k}}}=\frac{1}{\beta}\sum_{\nu_{n}}A_{l,{\bf{k}}}(\nu_{n})
=\sum_{i=1,2}\frac{(-1)^{l+i}}{2}\left[\frac{\eta_{{\bf{k}}}}{\sqrt{D_{\bf{k}}}}-(-1)^{l+i-1}\right]n_{\rm F}\left(\kappa_{i,{\bf{k}}}\right).
\label{Equation_35}
\end{eqnarray}
The plots of the functions $g_{l,{\bf{k}}}$ $l=1,2$ are presented in Figs.~\ref{fig:Fig_4} and ~\ref{fig:Fig_5}. In Fig.~\ref{fig:Fig_4}, the functions $g_{l,{\bf{k}}}$ are plotted for the case $W/|t_{1}|=0.75$ and $t_{\perp}=0.04|t_{1}|$. The temperature is set at $T/|t_{1}|=0.5$, and the intralayer Coulomb interaction parameter is fixed at $U/|t_{1}|=4$. The functions $g_{l,{\bf{k}}}$ in Figs.~\ref{fig:Fig_4} and ~\ref{fig:Fig_5} are plotted in the extended BZ scheme. We see clearly the hole layer sockets and the electron layer mounds in Fig.~\ref{fig:Fig_4}. In Fig.~\ref{fig:Fig_5}, the same functions are plotted for the case of higher temperature: $T/|t_{1}|=1.5$, and the considerable amplitude decrease is observable in Fig.~\ref{fig:Fig_5}.  

\begin{figure}
\begin{center}
\includegraphics[scale=.3]{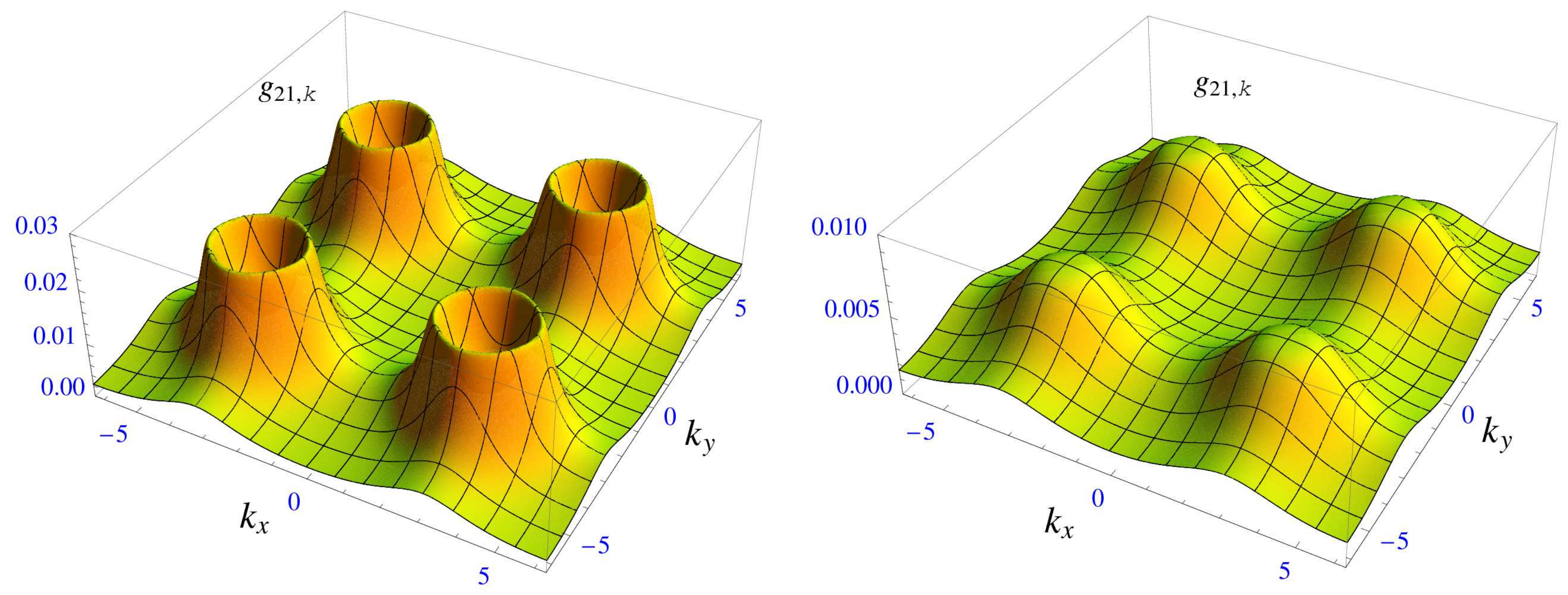}
\caption{\label{fig:Fig_6}(Color online)  The plot of the frequency-summed anomalous function $g_{21,{\bf{k}}}$ as a function of the ${\bf{k}}=k_{x}{\bf{G}}_{1}+k_{y}{\bf{G}}_{2}$ wave vector, in the extended BZ scheme. The left panel shows the ${\bf{k}}$ dependence at the fixed temperature $T/|t_{1}|=0.5$, and the right panel shows the ${\bf{k}}$ dependence at the fixed temperature $T/|t_{1}|=1.5$.}
\end{center}
\end{figure}

In Fig.~\ref{fig:Fig_6}, we have evaluated the anomalous, frequency-summed Green function $g_{21,{\bf{k}}}$, corresponding to the anomalous Green function $G_{21,{\bf{k}}}(\nu_{n})$. In the real space, the anomalous Green function is defined as  
%
%
\begin{eqnarray}
G^{\sigma,\sigma}_{21}(i\tau,i\tau)=\left\langle c^{\dag}_{2i,\sigma}(\tau)c_{1i,\sigma}(\tau)\right\rangle
\label{Equation_36}
\end{eqnarray}
In the Fourier representation we get 
\begin{eqnarray}
G^{\sigma,\sigma}_{21}(i\tau,i\tau)=\frac{1}{\beta{N}}\sum_{{\bf{k}},\nu_{n}}G^{\sigma\sigma}_{21,{\bf{k}}}(\nu_{n}),
\label{Equation_37}
\end{eqnarray}
where
\begin{eqnarray}
G^{\sigma\sigma}_{21,{\bf{k}}}(\nu_{n})=\frac{1}{\beta{N}}\left\langle c^{\dag}_{2{\bf{k}},\sigma}(\nu_{n})c_{1{\bf{k}},\sigma}(\nu_{n})\right\rangle.
\label{Equation_38}
\end{eqnarray}
Then, for the function $g_{21,{\bf{k}}}$, we have
\begin{eqnarray}
g_{21,{\bf{k}}}=-\frac{1}{\beta}\sum_{\nu_{n}}G_{21,{\bf{k}}}(\nu_{n})
=-\frac{t_{\perp}+\Delta}{2\sqrt{D_{\bf{k}}}}\left[n_{\rm F}(\kappa_{1,{\bf{k}}})-n_{\rm F}(\kappa_{2,{\bf{k}}})\right].
\label{Equation_39}
\end{eqnarray}
In the left panel in Fig.~\ref{fig:Fig_6}, we have presented the plot of the function in Eq.(\ref{Equation_39}) for the temperature fixed at $T/|t_{1}|=0.5$. The concave-like structure of the anomalous function $g_{21,{\bf{k}}}$ is the manifestation of the electron-hole pairing between different layers, and pair condensation in the DL system, which takes place at the same time for the given parameters ranges (as we will see later on, in the next Section \ref{sec:Section_5_2}).

The right panel, in Fig.~\ref{fig:Fig_6}, corresponds to the temperature fixed at $T/|t_{1}|=1.5$, and a considerable decrease of the $g_{21,{\bf{k}}}$ function amplitude is observable in the picture. The concave structure is also flattened in this case, due to the temperature effect.  
\renewcommand\thesubsection{\thesection.\arabic{subsection}}
%
\subsection{\label{sec:Section_5_2} Normal DOS functions}
%
Here, we calculate the ${\bf{k}}$-integrated, single-particle DOS functions $\rho_{l}(\nu)$ for different layers $l=1,2$. We obtain
\begin{eqnarray} 
\rho_{l}(\nu)=\sum_{\substack{i=1,2\\ j=1,2}}\frac{(-1)^{l+j-2}\rho_{\rm 2D}\left(\epsilon_{i}\right)}{2|\Lambda_{j}\left(\epsilon_{i}\right)|}\left[\frac{\eta\left(\epsilon_{i}\right)}{\sqrt{D\left(\epsilon_{i}\right)}}-(-1)^{l+j-1}\right].
\label{Equation_40}
\end{eqnarray}
%
\begin{figure}
\begin{center}
\includegraphics[width=320px,height=200px]{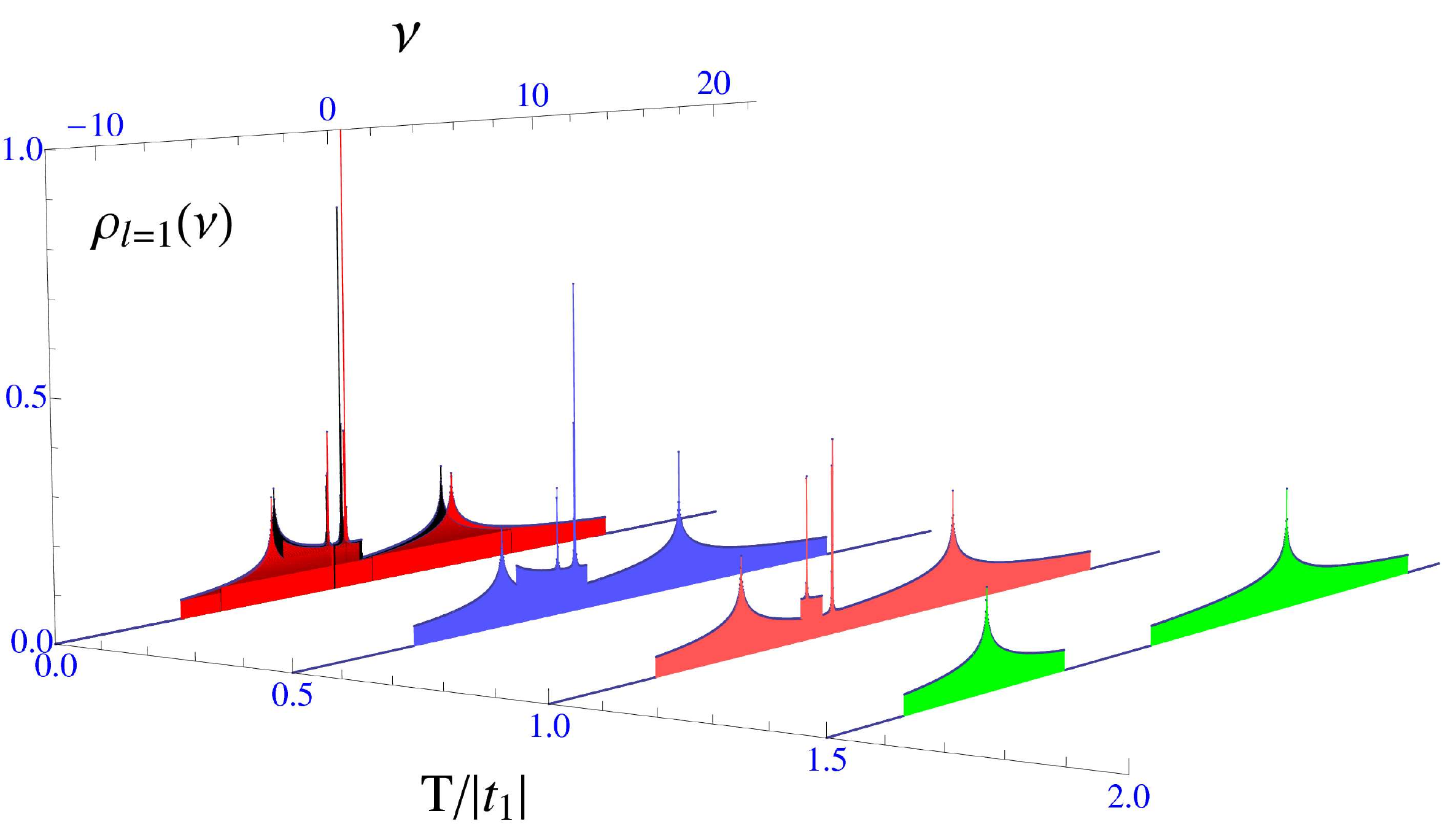}
\caption{\label{fig:Fig_7}(Color online) Normal DOS function $\rho_{l}(\nu)$ at $U/|t_{1}|=4$, for the hole layer with $l=1$, and at the weak interlayer coupling limit $W/|t_{1}|=0.75$.}
\end{center}
\end{figure} 
%
\begin{figure}
\begin{center}
\includegraphics[width=320px,height=200px]{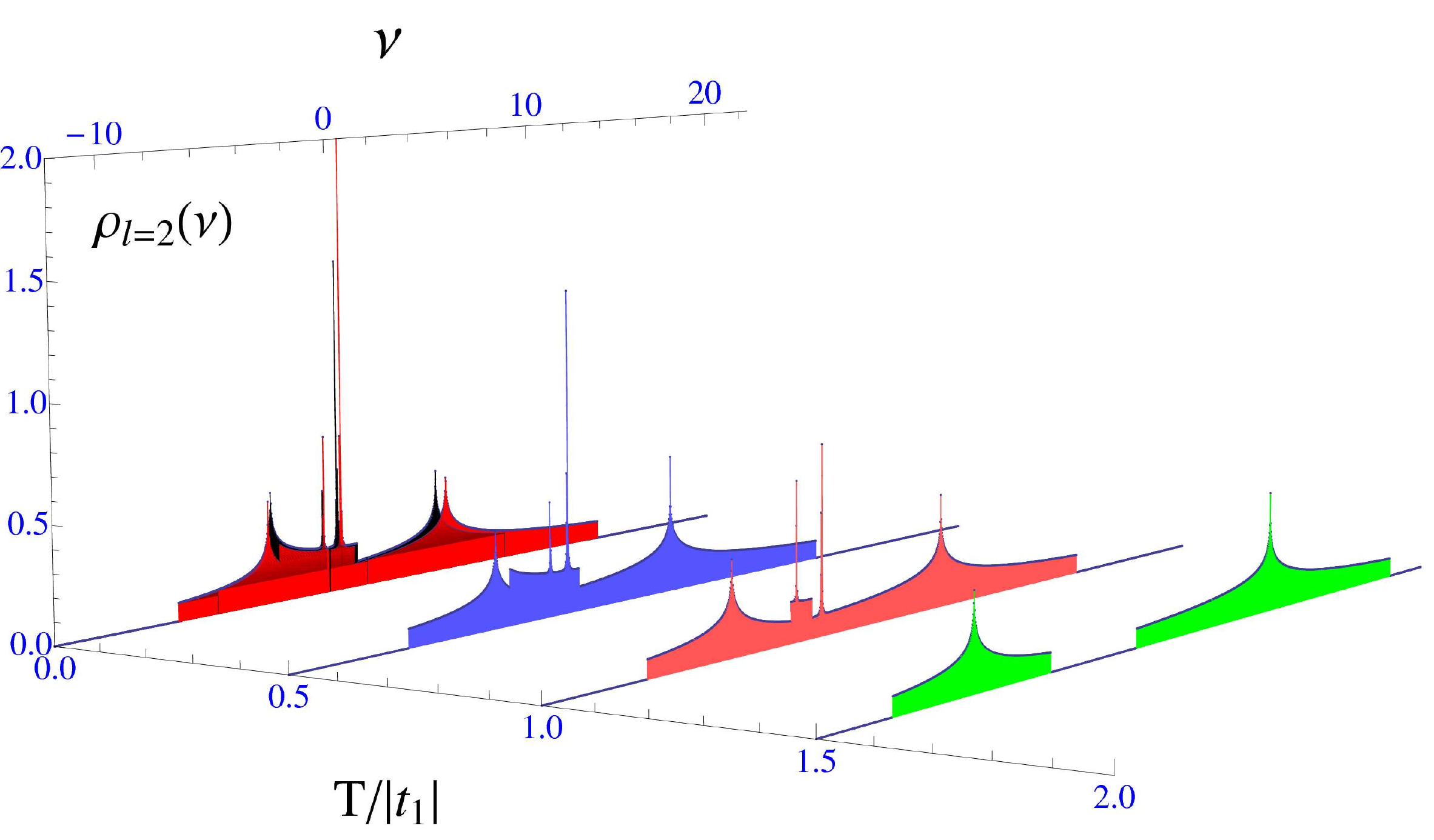}
\caption{\label{fig:Fig_8}(Color online) Normal DOS function $\rho_{l}(\nu)$ at $U/|t_{1}|=4$, for the electron layer with $l=2$, and at the weak interlayer coupling limit $W/|t_{1}|=0.75$.}
\end{center}
\end{figure} 

The derivation of this result is given in details in the Appendix \ref{Section_8}. We have introduced in Eq.(\ref{Equation_40}) the non-interacting elliptic DOS function for the 2D square lattice 
\begin{eqnarray}
\rho_{\rm 2D}(z)=\frac{1}{\pi^{2}}K\left(\sqrt{1-\frac{z^{2}}{4}}\right),
\label{Equation_41}
\end{eqnarray}
where $K(z)$ is the complete elliptic integral of the first kind \cite{cite-45}. The functions $\eta(z)$, $D(z)$, and $\Lambda_{j}\left(z\right)$ $j=1,2$, in Eq.(\ref{Equation_40}), are defined in the following way:
\begin{eqnarray}
\eta(z)=2(t_{1}-t_{2})z-\mu,
\label{Equation_42}
\newline\\
D(z)=\eta^{2}(z)+(t_\perp+\Delta)^{2},
\label{Equation_43}
\newline\\
\Lambda_{j}(z)=2(t_{1}+t_{2})+(-1)^{j-1}\frac{\eta(z)\eta'(z)}{\sqrt{D(z)}}.
\label{Equation_44}
\end{eqnarray}
Here $\eta'(z)$ means the derivative of the function $\eta(z)$. The parameters $\epsilon_{i}$ $i=1,2$, in the arguments of the functions in Eq.(\ref{Equation_40}) are defined in the following way
\begin{eqnarray}
\epsilon_{i}=-\frac{W_{1}+(-1)^{i}\sqrt{|W_{2}|^{2}+4t_{1}t_{2}\left(t_{\perp}+\Delta\right)^{2}}}{8t_{1}t_{2}},
\label{Equation_45}
\end{eqnarray}
where the functions $W_{i}$ $i=1,2$ are 
\begin{eqnarray}
W_{i}=\mu\left[t_{1}+(-1)^{i}t_{2}\right]+\left[t_{1}+(-1)^{i-1}t_{2}\right]\left(W-\nu-\frac{U}{2}\right).
\label{Equation_46}
\end{eqnarray}

Furthermore, we have evaluated the DOS functions given in Eq.(\ref{Equation_40}) (for both layers $l=1,2$) using the highest precision accuracy for the elliptic function given in Eq.(\ref{Equation_41}). 

In Fig.~\ref{fig:Fig_7} we have shown the temperature dependence of the normal DOS function for the layer-1 holes, for the fixed value of the intralayer Coulomb interaction parameter $U/|t_{1}|=4$, and for the fixed value of the interlayer hopping amplitude $t_{\perp}=0.04|t_{1}|$. We see in Fig.~\ref{fig:Fig_7} that the DOS structure is spreading well, along the $\nu$-axis, when increasing the temperature. 

\begin{figure}
\begin{center}
\includegraphics[scale=.57]{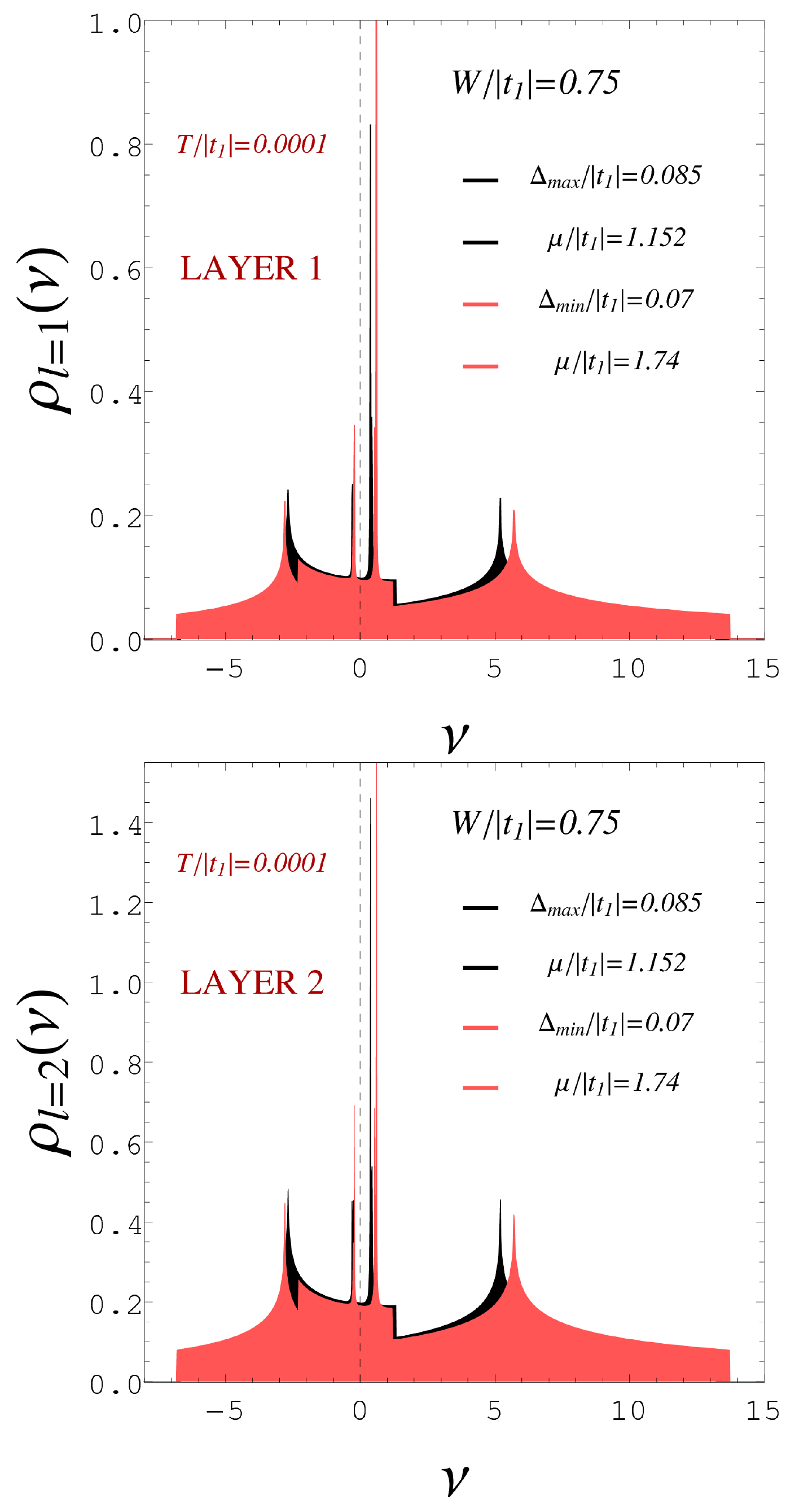}
\caption{\label{fig:Fig_9}(Color online) Normal DOS function $\rho_{l}(\nu)$ at $U/|t_{1}|=4$, for the hole layer with $l=1$ (see the upper panel) and electron layer with $l=2$ (see the lower panel). The temperature is set at $T/|t_{1}|=0.0001$ and the small interlayer coupling limit is considered $W/|t_{1}|=0.75$.}
\end{center}
\end{figure}

In Fig.~\ref{fig:Fig_8} we have shown the DOS structure for the layer-2 electrons. A relatively small interlayer interaction limit $W/|t_{1}|=0.75$ is considered in Figs.~\ref{fig:Fig_7} and ~\ref{fig:Fig_8}. We see in Figs.~\ref{fig:Fig_7} and ~\ref{fig:Fig_8} that at the low-temperature limit (i.e., when $0.0001\leq T/|t_{1}|\leq 1.0$), the DOS structures for both layers $l=1,2$ are gapless, which is the manifestation of the excitonic condensates regime \cite{cite-28, cite-29} in the DL system. In the relatively high-temperature limit, (see the DOS structures at $T=1.5$, in Figs.~\ref{fig:Fig_7} and ~\ref{fig:Fig_8}) a very large hybridization gap is opening, which signals the passage to the BI limit, without excitonic condensates. It is interesting to discuss separately the DOS shapes at $T/|t_{1}|=0.0001$ in Figs.~\ref{fig:Fig_7} and ~\ref{fig:Fig_8}, because of its degenerate character. In fact, there are many DOS solutions, in this case, as it is presented in Fig.~\ref{fig:Fig_9}, and this is due to the complicated behavior of the excitonic gap parameter and chemical potential at the very low-temperature limit (see in the Section \ref{sec:Section_4} and also in Ref.\onlinecite{cite-33}). Indeed, the excitonic gap parameter, for the case $T/|t_{1}|=0.0001$, small interlayer coupling parameter $W/|t_{1}|=0.75$, and small interlayer hopping amplitude $t_{\perp}=0.04|t_{1}|$, has many solutions corresponding the same value of the intraplane Coulomb interaction parameter $U/|t_{1}|$. Namely, there are 16 solutions for $\Delta$ and $\mu$ at $T/|t_{1}|=0.0001$ and at $U/|t_{1}|=4$ (see in Fig.~\ref{fig:Fig_1} and also in Ref.\onlinecite{cite-33}). Particularly, in Fig.~\ref{fig:Fig_9}, we have presented the normal Frenkel DOS structures corresponding to the maximal (see the black area in Fig.~\ref{fig:Fig_9}) and minimal  (see the area in red, in Fig.~\ref{fig:Fig_9}) values of the excitonic pairing gap parameter at $U/|t_{1}|=4$. The corresponding values of the chemical potentials have been calculated and are given in Fig.~\ref{fig:Fig_9}. We see, in Fig.~\ref{fig:Fig_9}, that the DOS structures for the maximal value of the gap parameter are slightly higher than for the case corresponding the minimum of the Frenkel channel gap parameter and also they are slightly displaced on the $\nu$-axis. This is true for both layers $l=1,2$. Actually, as it is indicated above, there are 14 more DOS structures, apart of presented two in Fig.~\ref{fig:Fig_9}, but we do not present here all of them.
%
%
Another important observation that could be gained from the DOS structures presented in Figs.~\ref{fig:Fig_7} and ~\ref{fig:Fig_8}, is related to the DOS peaks positions on the frequency axis $\nu$. Indeed, in this case, we see clearly that the excitonic condensates double peaks are present in the DOS spectra (see the well pronounced double peaks in between the outermost excitonic single-particle excitation peaks in Figs.~\ref{fig:Fig_7} and ~\ref{fig:Fig_8}). Both, condensates, and single-particle excitation peaks are decreasing when augmenting the temperature. The important feature that should be mentioned here is that the condensates double peak structure disappears at the very high-temperature limit (see the DOS structures for the case $T/|t_{1}|=1.5$), while the single-particle excitation peaks remain present in the DOS spectra with a well-defined hybridization gap, which is signaling the passage to the BI phase in the system.  

\begin{figure}
\begin{center}
\includegraphics[width=320px,height=200px]{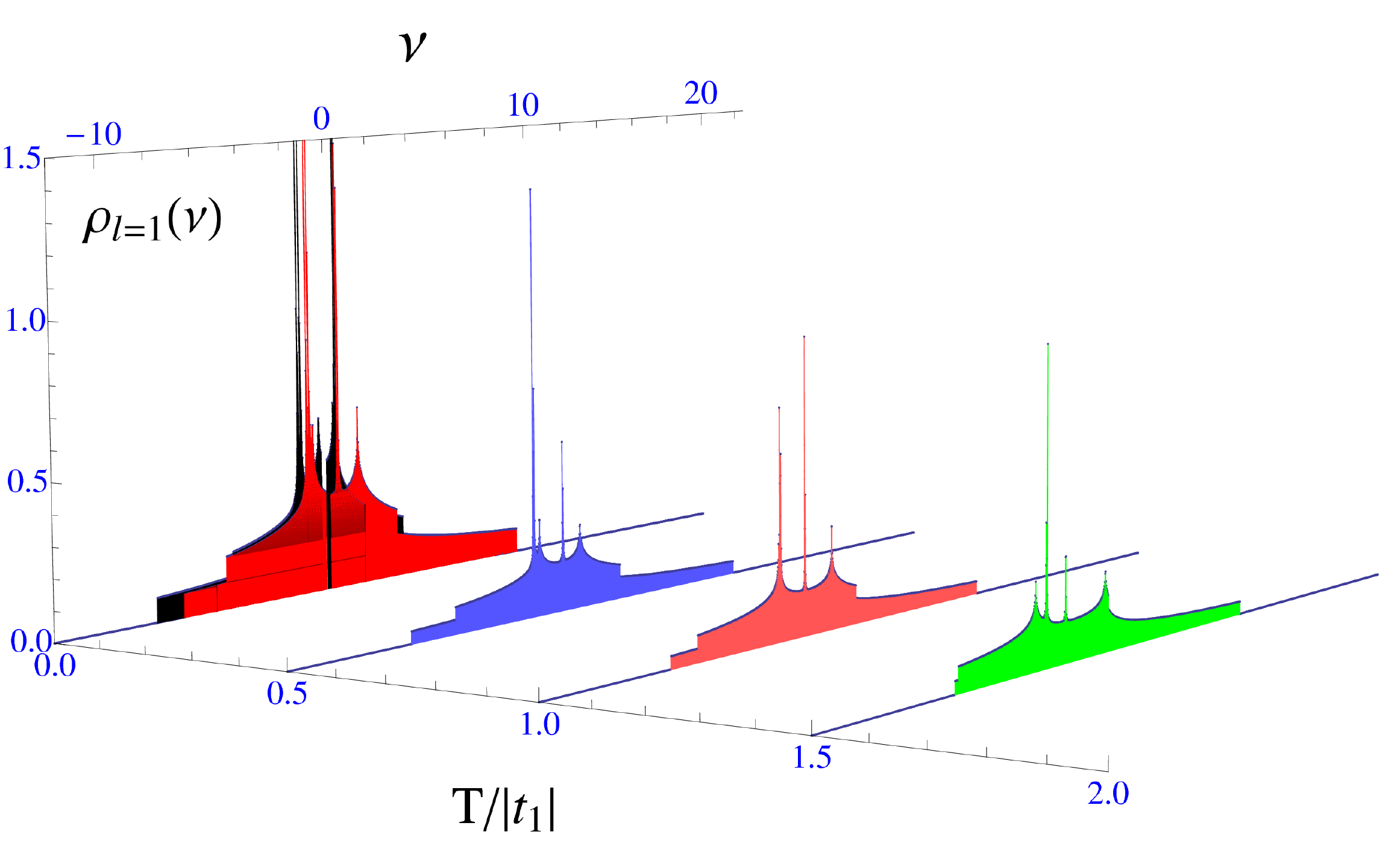}
\caption{\label{fig:Fig_10}(Color online)  Normal DOS function $\rho_{l}(\nu)$ at $U/|t_{1}|=4$, for the hole layer with $l=1$, and at the strong interlayer coupling limit $W/|t_{1}|=1.5$.}
\end{center}
\end{figure}

\begin{figure}
\begin{center}
\includegraphics[width=320px,height=200px]{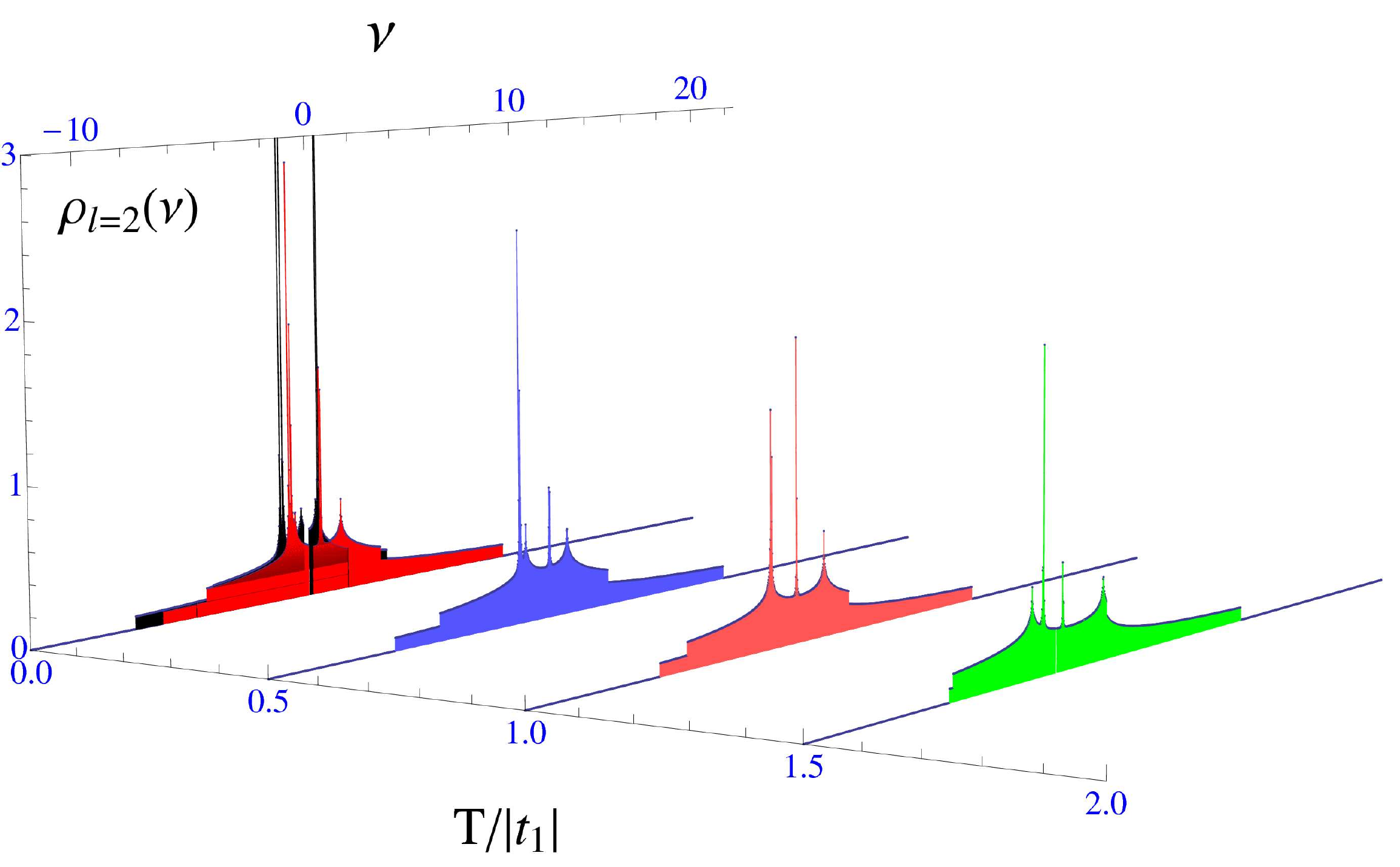}
\caption{\label{fig:Fig_11}(Color online) Normal DOS function $\rho_{l}(\nu)$ for the electron layer with $l=2$, and at the strong interlayer coupling limit $W/|t_{1}|=1.5$.}
\end{center}
\end{figure}

\begin{figure}
\begin{center}
\includegraphics[scale=.57]{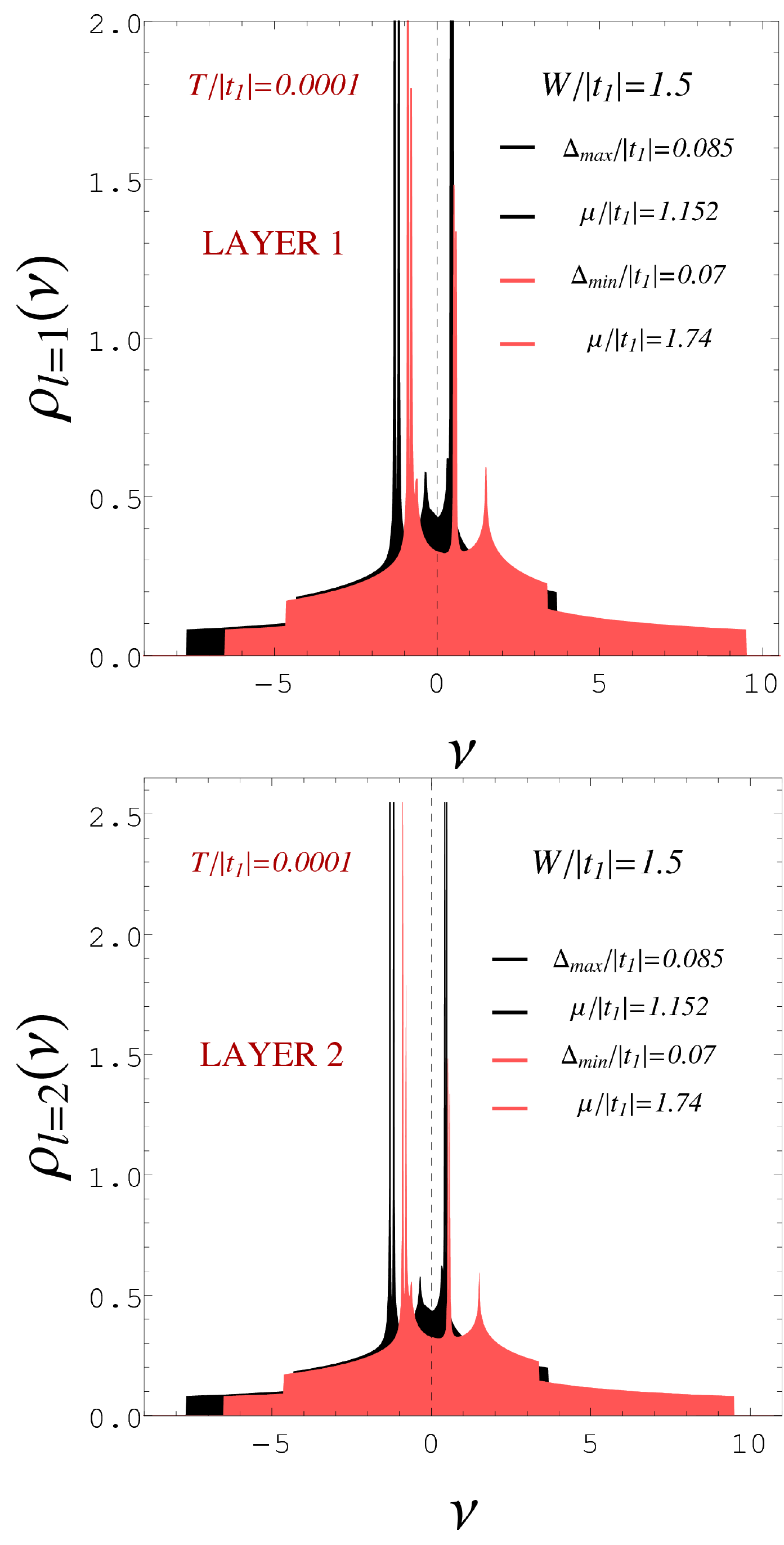}
\caption{\label{fig:Fig_12}(Color online) Normal DOS function $\rho_{l}(\nu)$ at $U/|t_{1}|=4$, for the hole layer with $l=1$ (see the upper panel) and electron layer with $l=2$ (see the lower panel). The temperature is set at $T/|t_{1}|=0.0001$, and the strong interlayer coupling limit is considered $W/|t_{1}|=1.5$.}
\end{center}
\end{figure}

In Figs.~\ref{fig:Fig_10} and ~\ref{fig:Fig_11}, we have presented the DOS results for the strong interlayer Coulomb interaction limit: $W/|t_{1}|=1.5$ and for $t_{\perp}=0.04|t_{1}|$. We see, that in this case the DOS amplitudes are much higher and the hybridization gap is totally absent (see the temperature evolution of DOS functions given in Figs.~\ref{fig:Fig_10} and ~\ref{fig:Fig_11}), for all temperature ranges. Thus, the DL system is always in the condensates regime in this case. Obviously, augmenting the interlayer Coulomb interaction favors the condensates order in the DL system, and the excitonic condensates are amplified for both layers $l=1,2$. Therefore, the parameter $W/|t_{1}|$ plays an important role in the physics of the DL structures. Mainly, by varying it, we can obtain the stable condensates states, even at the higher temperatures. We see also in Figs.~\ref{fig:Fig_10} and ~\ref{fig:Fig_11}, that the excitonic condensates peaks are strongly correlated with the excitonic excitation peaks, in this case, and they are more difficult to distinguish than in the previous case. This result is related to the strong quantum coherence effects in the DL system in the case of strong interlayer coupling, which is not the case for the smaller values of the parameter $W/|t_{1}|$ (see in Figs.~\ref{fig:Fig_7} and ~\ref{fig:Fig_8})). But the general behavior for the positions on the $\nu$-axis is the same also for this case, and the condensates double peaks are situated between the single-particle excitation outermost peaks.
%
%
%
In Fig.~\ref{fig:Fig_12} we have presented the DOS structures  at $T/|t_{1}|=0.0001$ (for the case $W/|t_{1}|=1.5$), which are given also in Figs.~\ref{fig:Fig_10} and ~\ref{fig:Fig_11} (see the first DOS structures in Figs.~\ref{fig:Fig_10} and ~\ref{fig:Fig_11}). The plots in Fig.~\ref{fig:Fig_12} correspond to the maximal (see the area in black in both panels in Fig.~\ref{fig:Fig_12}) and minimal  (see the area in red in both panels in Fig.~\ref{fig:Fig_12}) values of the excitonic pairing gap parameter at $U/|t_{1}|=4$. The corresponding values of the chemical potentials have been calculated and are given in Fig.~\ref{fig:Fig_12}. We see again, in Fig.~\ref{fig:Fig_12} that the DOS structures for the maximal value of the gap parameter are slightly higher than for the case corresponding the minimum of the gap parameter and also they are displaced on the $\nu$-axis. Actually, there are 20 more DOS structures apart of presented two in Fig.~\ref{fig:Fig_12}, due to the complicated pairing gap behavior at low-temperature case (see in Ref.\onlinecite{cite-33}, where there are 22 solutions for the gap parameter, corresponding to the same value of the intralayer Coulomb interaction parameter $U/|t_{1}|=4$).
%
%
%
Thus the passage into the BI phase is only possible for the small values of the interlayer Coulomb interaction parameter and at small interlayer hopping amplitudes (as $t_{\perp}=0.04|t_{1}|$, in our case). Thus, by varying the interlayer interaction parameter, gives the possibility to define the conditions, when different phases could be present in the DL system. At the small interlayer coupling regime, and at the low-temperature case, we have the coexistence of two different phases of matter, namely the excitonic condensates with the preformed excitonic pairs and the excitonic pair formation phases (or the excitonic insulator phase \cite{cite-28, cite-29, cite-30, cite-36, cite-37, cite-38, cite-39, cite-40}). For the high-temperature limit, the passage from the excitonic BEC to the BI phase is continuously driven by the temperature effect. Contrary, for the strong interlayer coupling regime, we have the coexistence of these two different phases of matter in all temperature ranges, and the excitonic condensates are present for the high-temperature limit as well. This effect is purposeful, regarding the efforts to obtain the room temperature condensates in the DL systems.   

\begin{figure}
\begin{center}
\includegraphics[scale=.52]{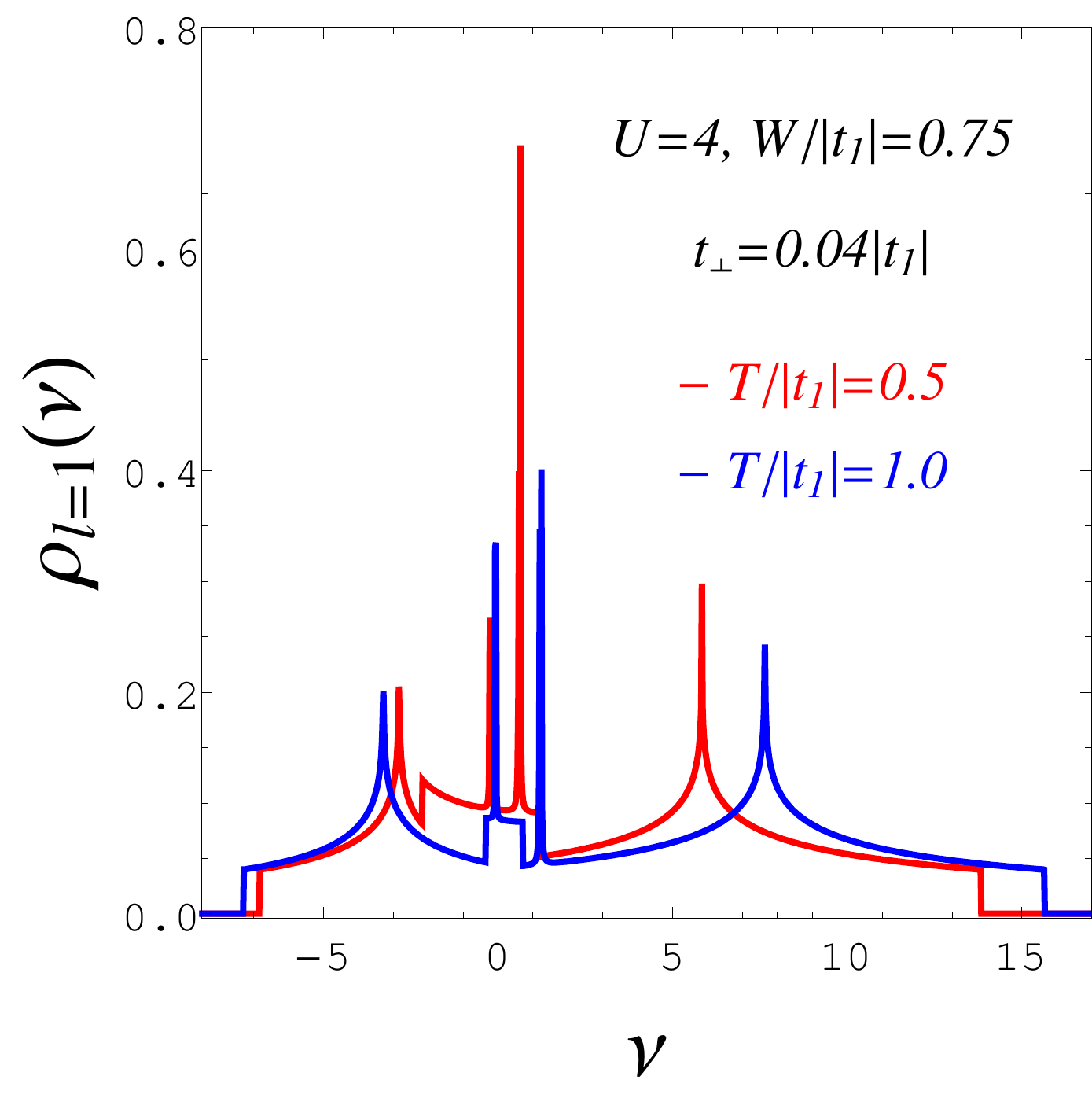}
\caption{\label{fig:Fig_13}(Color online) online) Normal DOS function $\rho_{l}(\nu)$ for the
hole layer with $l=1$. The figure shows the displacement of the condensate and single-particle excitation peaks positions on the frequency $\nu$-axis, when increasing the temperature.}
\end{center}
\end{figure}

We see also in Figs.~\ref{fig:Fig_10} -~\ref{fig:Fig_11}, that when increasing the interlayer interaction parameter $W/|t_{1}|$, the DOS amplitudes become considerably higher, and when increasing the temperature, the positions of both: condensates and single-particle excitation peaks are far and away on the $\nu$-axis and this is shown in Fig.~\ref{fig:Fig_13} for the case of the hole layer with $l=1$ and at $W/|t_{1}|=0.75$.

\begin{figure}
\begin{center}
\includegraphics[width=320px,height=200px]{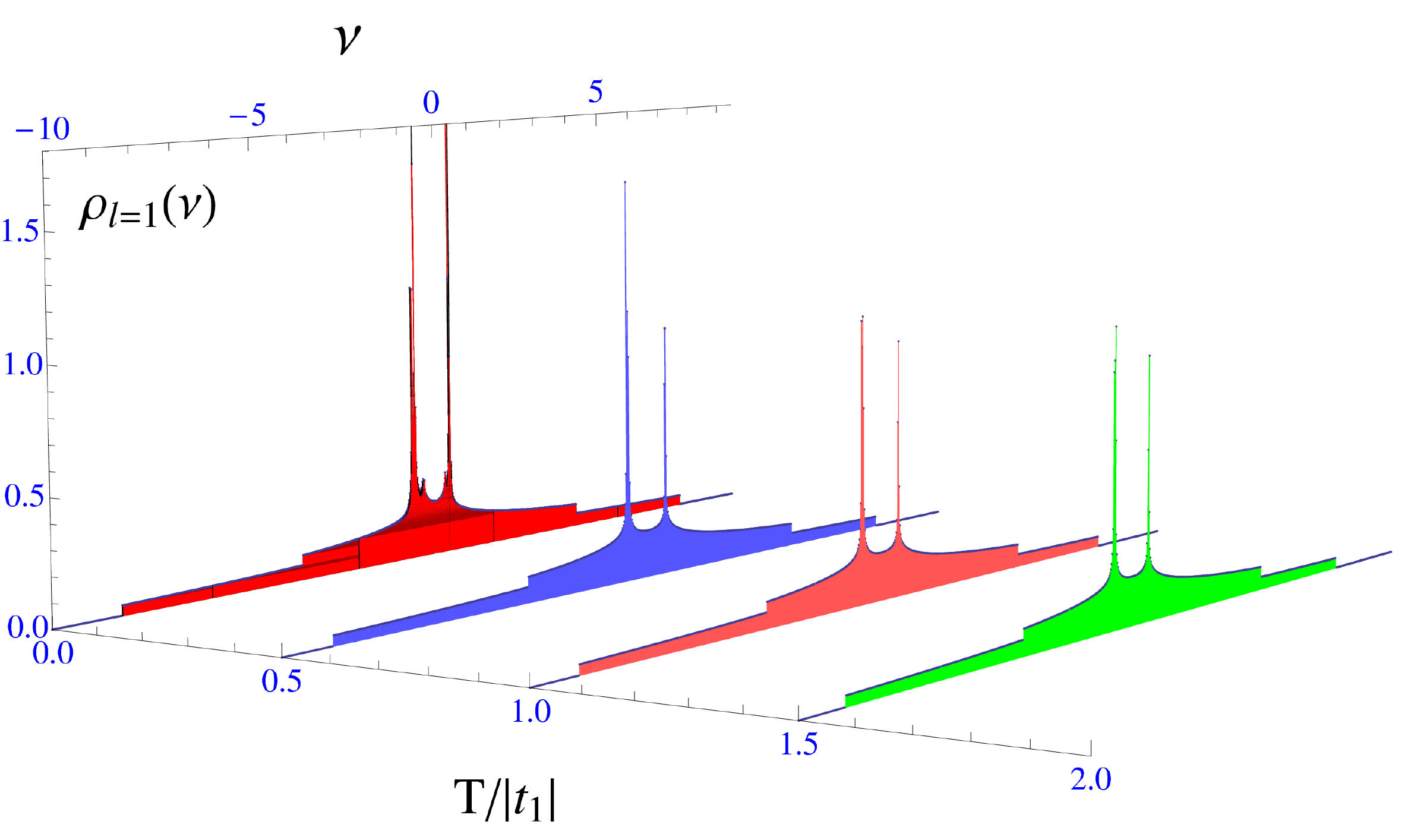}
\caption{\label{fig:Fig_14}(Color online) online)Normal DOS function $\rho_{l}(\nu)$ at the stable bunching point $U_{B}/|t_{1}|=1.5$, for the hole layer with $l=1$. The interlayer Coulomb interaction parameter is set at $W/|t_{1}|=0.75$, and $t_{\perp}=0.04|t_{1}|$.}
\end{center}
\end{figure}

\begin{figure}
\begin{center}
\includegraphics[width=320px,height=200px]{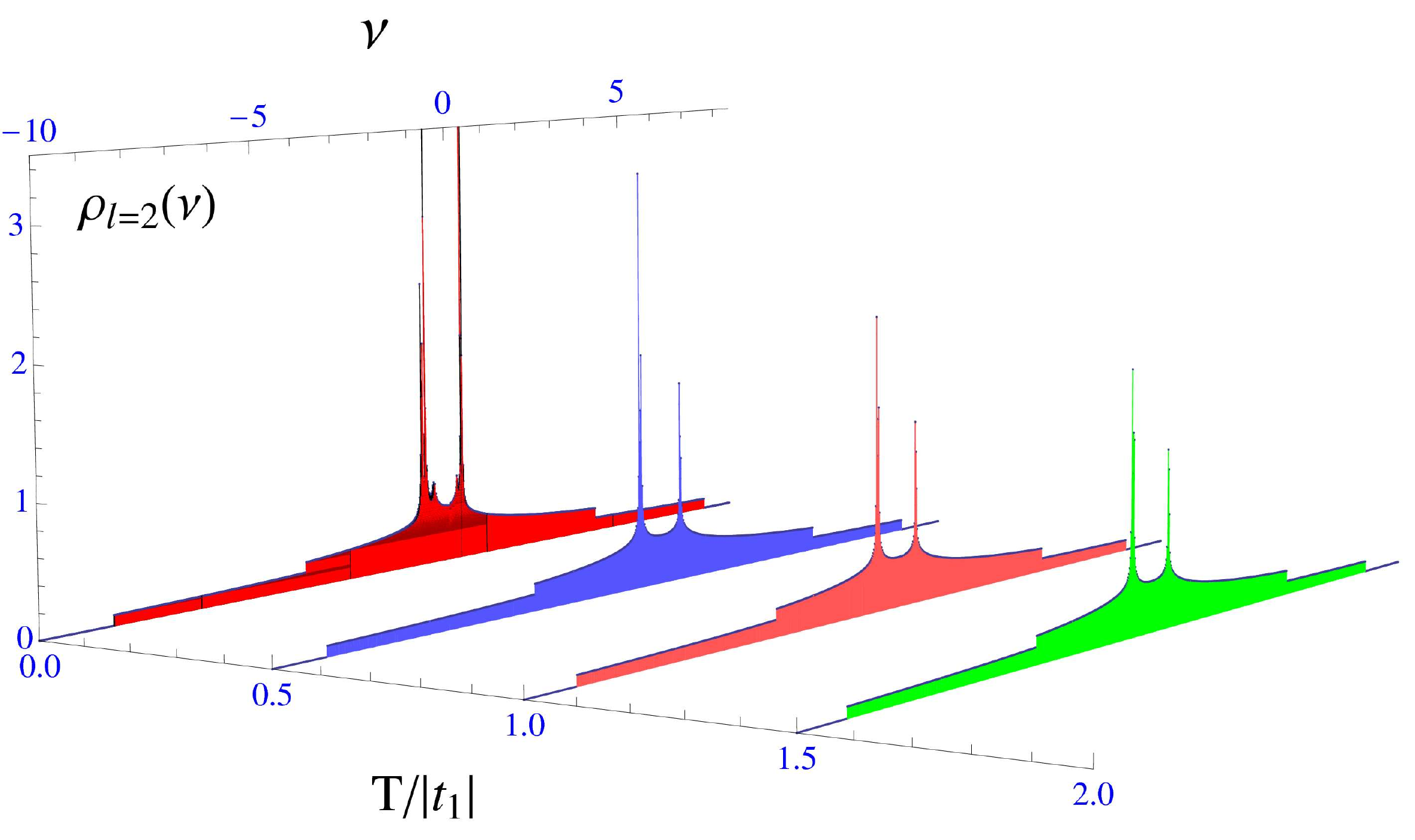}
\caption{\label{fig:Fig_15}(Color online) Normal DOS function $\rho_{l}(\nu)$ at the stable bunching point $U_{B}/|t_{1}|=1.5$, for the electron layer with $l=2$. The interlayer Coulomb interaction parameter is set at $W/|t_{1}|=0.75$, and $t_{\perp}=0.04|t_{1}|$.}
\end{center}
\end{figure}

The situation is more drastic at the stable bunching point $U_{B}/|t_{1}|$, which remains unchanged when increasing the temperature (see the discussion in the Section \ref{sec:Section_5_1} and also in Ref.\onlinecite{cite-33}). In Figs.~\ref{fig:Fig_14} and ~\ref{fig:Fig_15}, we have presented the DOS structures at $U_{B}/|t_{1}|=1.5$ corresponding to the values $W/|t_{1}|=0.75$ and $t_{\perp}=0.04|t_{1}|$ (see the stable bunching point solution at $U_{B}/|t_{1}|=1.5$ in Fig.~\ref{fig:Fig_1}, in the Section \ref{sec:Section_4}, and also in Fig.~\ref{fig:Fig_3}, in the Section \ref{sec:Section_5_1}). 
%
%
%
%
%
In this case, the DOS structures are also gapless, and the excitonic condensates peaks are coinciding exactly with the single-particle excitation ones, thus increasing the coherence effects in the DL system. In addition, the DOS structures at $U_{B}/|t_{1}|$, still unchanged, when augmenting the temperature, and, only a relatively small amplitude decrease of DOS double peaks amplitudes is observable. This is consistent with the definition of the point $U_{B}/|t_{1}|$ (see in Fig.~\ref{fig:Fig_1}, in the Section \ref{sec:Section_4}). Here, the DOS structure is also degenerated at the very low-temperature $T/|t_{1}|=0.0001$, due to the degenerate gap structure at $U/|t_{1}|=1.5$ (see the discussion in Ref.\onlinecite{cite-33}). There are 18 solutions for the gap parameter corresponding to the value $U/|t_{1}|=1.5$ \cite{cite-33}. The difference with the previous cases is that the DOS spectra are exactly coinciding at the bunching point, and there are not displacements in the peaks positions for different DOS solutions. We see in Figs.~\ref{fig:Fig_14} and ~\ref{fig:Fig_15} that the condensates orders are strongly enhanced at the stable bunching point $U_{B}$, and the DOS amplitudes are considerably higher than for the cases away from the point $U_{B}$. In other words, this point is related to the stable condensates sources in the system, for all temperature ranges.         

We observe also in all Figs.~\ref{fig:Fig_7} -~\ref{fig:Fig_15}, that the DOS amplitudes are significantly higher for the electron layer-2, with $l=2$ normal DOS spectra. 
%
\section{\label{sec:Section_6} Discussion}
%
Parallelizing with the results of the works given in Refs.\onlinecite{cite-27, cite-28, cite-29, cite-33} we can conclude that the DL system is characterized by the very strong electron-hole correlations, which are responsible for the excitonic condensation in DL. Especially, the gapless behavior in the excitonic normal DOS structure is a perfect manifestation of the condensates artifact in the DL structures. Even more detailed phase coherence analysis, that could be done as the next step of the presented theory, will not affect considerably this result. We expect that only the minor corrections will appear on the mechanism of the excitonic condensates states formation presented here. The treatment of the phase sector in the problem would be important in order to distinct the interlayer superfluid state formation phase \cite{cite-25} from the excitonic condensate state discussed here. This could be done by taking into account the role of the particle phase coherence and by considering the vortex-antivortex bound phase coherent states \cite{cite-30}. 

From the experimental side, the coherent light emission measurements will be insufficient in this case and taking into account the possible laser coherent emission interference effects. This is in contrast with the polariton condensates \cite{cite-46, cite-47, cite-48}, where the intraplane coherence is directly measured by the first-order light emission interference. For the indirect excitonic condensates, such in our case, the first-order interference measurements imply requires the additional experimental refinements, in order to distinguish the excitonic coherence spectral lines from the standard coherent lasing effects. However, it is not a good test to demonstrate the coherent excitonic condensates, because of the possible exciton dephasing due to the strong interlayer correlations. Meanwhile, the mentioned techniques could be used for measuring the spontaneous hybridization in the system, for the case of the small interlayer Coulomb interaction and high-temperature limit. As it is shown above, in this case, the system is in the BI phase in the crossover scenario, thus the macroscopic phase coherent state is absent in this case. 
        
Another test to prove the BI state in the system could be proposed with the help of the photoluminescence (PL) measurements. Indeed, in the experimental light absorption measurements, the photoexcited electrons, with the sufficient excitation energy in the layer-1 jump into the layer-2, leading the exciton formation across the DL structure. At sufficiently low temperatures, the absorption spectrum is sufficiently broad, in order to neglect the electron recombination effects in the layer-2. But for the higher temperature and,   
as it follows from the discussion above, the experimental evidence of the BI state of excitons in the correlated DL structure could be signaled only by the spontaneous hybridization between the valence and conduction bands for the given parameters ranges (for the small interlayer coupling limit, and high temperatures). Therefore, as the experimentally accessible quantities in the photoluminescence (PL) measurements, we have the normal DOS functions presented in Figs.~\ref{fig:Fig_7} -~\ref{fig:Fig_9}). However, in order to measure the condensates parts of the DOS, corresponding the lower temperatures, the only hybridization measurements are not sufficient. 

Another difficulty, related to the extraction of the coherent exciton DOS from only the PL line shapes (with a fixed photon energy), is due to the origin of the Stokes shift, and PL line broadening \cite{cite-49}. The measurement by the PL technique of the real absorption for the normal coherent DOS is problematic, also because of the strong scattering effects, which are destabilizing the condensates states. 

In fact, the direct experimental measurement of the coherent excitonic DOS is rather a difficult task, which asks highly refined experimental techniques. We suggest that the normal DOS functions, given in Eq.\ref{Equation_40}, and for all parameter ranges, could be experimentally verified by analyzing of the transmittance and photoluminescence excitation (PLE) spectra \cite{cite-50}. Using the optics relations,
one can obtain the absorption spectrum from the measured transmittance one. Furthermore,
the analysis of the absorption spectra will provide a way to determine the partial excitonic DOS. Here, a special attention has to be paid when preparing the sample for the measurements. For the DOS spectra modeling procedure, the fitting parameters have to be taken into account carefully, such, like the width of the sample size distribution (which is usually taken as Gaussian), the broadening parameter, and the thickness of the film \cite{cite-51}. 

In contrary to PL measurements, in the conventional PLE technique, the fixed photon energy is changed into the excitation energy and PLE spectroscopy could be an alternative solution of that problem, assuming that the photoexcited e-h pairs always end-up at the lowest energy states (in our case, being simultaneously in the BEC states). The spectrum measured by the PLE techniques is strongly correlated to the absorption spectrum of the sample.

Alternatively, in order to measure the excitonic DOS, given in  Eq.(\ref{Equation_40}), we suggest
also the sensible techniques of integrated photoluminescence excitation (IPLE). The direct measurement of the coherent excitonic DOS by PLE techniques, mentioned above, could be difficult in some cases, related to the thickness of the samples and significant extinction caused by the DL films inhomogeneities. The IPLE techniques these difficulties are avoided, and the IPLE spectrum is arising from the integration of the excitonic PL lineshape, which gives a very good estimate for the shapes of the absorption spectrum, taking into account also the absorption spectra for higher excited energies.   
%
\section{\label{sec:Section_7} Conclusion}
%
In the present paper, we considered the excitonic gap formation and excitonic DOS in the double layer electronic structure, composed of holes (in the layer-1) and electrons (in the layer-2). Employing a pertinent to this case theoretical formalism, we have obtained a system of coupled self-consistent nonlinear equations for the excitonic gap parameter and chemical potential in the system at half-filling. We have considered the case of the zero external electric field, and the hopping asymmetric DL structure: $|t_{1}|\neq |t_{2}|$. From the numerical results for the excitonic gap parameter, we have shown that the strong and weakly bound excitonic formations are possible in the correlated DL system and we have found a stable bunching point $U_{B}$ in the spectrum of the chemical potential. 

Furthermore, we have calculated the excitonic normal DOS functions for different interlayer Coulomb interaction regimes. In the case of small interlayer coupling, we have found a gapless DOS behavior for the case of low temperatures, while in the high-temperature region, the DOS structure exhibit a very large, spontaneous hybridization gap, and destruction of the condensate order is enhanced. From the structure of the excitonic normal DOS functions, we have shown a strong evidence of the exciton condensation in the DL system. These condensates states disappear at high-temperature and small interlayer interaction regime, which could be described as a smooth passage from the coherent BEC regime into the uncorrelated excitonic pairs. In contrast, for the strong interlayer coupling limit, the condensates states persist in the high-temperature limit and the system is always in the BEC regime in this case. 

Regarding the previous results on this subject of the exciton condensation in the 3D systems with phase coherence mechanism \cite{cite-27, cite-28, cite-29}, it is especially important to mention here, that the DL structure itself, amplifies the intensity of the exciton condensation, without additional phase coherence mechanisms, or applied external fields. This is very important from the theoretical point of view and is crucial when considering the exciton condensation mechanisms. Thus, the general conclusion to be gained from our considerations is that the excitonic pair formation and their condensation into the fundamental state occurs simultaneously in the DL structures, which is in contrast to the 3D bulk case \cite{cite-28, cite-29}, where the excitonic pair formation and excitonic condensation are two distinct phases of matter, which emerge at different temperature scales \cite{cite-27,cite-28,cite-29,cite-30,cite-31,cite-32}. 

Since the electron-hole pairs in the excitonic DL systems couple directly to the emitted photos, and since the intraplane coherence maps well with the optical coherence of the emitted photons, then, as the future investigation of our results obtained here, we will calculate the intraplane and interplane coherence lengths. For the interplane correlations, two coherence lengths should be clearly distinguished: one related to the Frenkel type excitons and the other with the WM excitonic formations, which are both possible in the DL structure \cite{cite-33}. This will give also another possibility (more simpler than the direct experimental measurements of the normal DOS spectra presented in the Section \ref{sec:Section_6}) to compare, in a best way, the results of our theory with the coherent light emission experiments, which not eat give the clear results about the exciton condensation in the DL systems.   

Nevertheless, in order to clarify the mechanism of the interlayer superfluid phase formation \cite{cite-25}, we have to properly include the particle coherence effects, by introducing the phase variables and by discussing the exciton pairing symmetry breaking, when lowering the temperature. We expect that the role of the electron phases \cite{cite-28, cite-29} would be important for the case of small interlayer interaction and high-temperature region. The observation of the stable condensates states, at the high-temperature and strong interlayer coupling regime, opens the new perspectives to observe the room-temperature exciton condensates in the DL structures. Alternatively, for the small interlayer coupling regime, there exists  a crossover mechanism from BEC to BI state.  

Another perspective of possible elaboration of the presented here theory is the inclusion of the biexcitonic effects in the problem via the quantum rotor approach and by integrating out the fermionic degrees of freedom. The topics of high importance would be the stability conditions for biexcitonic formations and their influences on the excitonic condensates formations in different parameter regimes. Due to the very short lifetime of the biexcitonic formations, we expect that their influences on the effects discussed here will not be much important.      
\appendix
%
\section{\label{Section_8} Single-particle DOS}
%
We will derive here the expression of the single-particle DOS functions $\rho_{l}(\nu)$ given in Eq.(\ref{Equation_40}), in the Section \ref{sec:Section_5_2}. We start by the definition of the normal single-particle DOS for the layers $l=1,2$, and we fix the spin variable $\sigma=\uparrow$. 
The single-particle Green function for the layer $l$ is defined as 
\begin{eqnarray}
G^{\uparrow, \uparrow}_{l}(i{\tau},i{\tau})=\frac{1}{\beta{N}}\sum_{{\bf{k}},\nu_{n}}G^{\uparrow, \uparrow}_{l,{\bf{k}}}(\nu_{n}),
\label{Equation_A_1}
\end{eqnarray}
where
\begin{eqnarray}
G^{\uparrow, \uparrow}_{l,{\bf{k}}}(\nu_{n})=-\frac{1}{\beta{N}}\left\langle c^{\dag}_{l{\bf{k}},\uparrow}(\nu_{n})c_{l{\bf{k}},\uparrow}(\nu_{n})\right\rangle.
\label{Equation_A_2}
\end{eqnarray}
Then, we use the expression of the inverse Green function matrix given in Eq.(\ref{Equation_17}), in the Section \ref{sec:Section_3}, and we derive the explicite expression for the single-particle fermion average $\left\langle c^{\dag}_{l{\bf{k}},\uparrow}(\nu_{n})c_{l{\bf{k}},\uparrow}(\nu_{n})\right\rangle$ using the functional derivation techniques \cite{cite-34}. The procedure is rather standard \cite{cite-28, cite-34}. We introduce the source field spinors $J_{{\bf{k}}}(\nu_{n})$ and $J^{\dag}_{{\bf{k}}}(\nu_{n})$ into the partition function given in Eq.(\ref{Equation_21}). They are sources, coupled to the Nambu spinor vectors, defined in the Section \ref{sec:Section_3}, in Eq.(\ref{Equation_16}). We have
\begin{eqnarray}
J_{{\bf{k}}}(\nu_{n})=\left[
J_{1{\bf{k}},\uparrow}(\nu_{n}), J^{\dag}_{1{\bf{k}},\downarrow}(\nu_{n}), J_{2{\bf{k}},\downarrow}(\nu_{n}), J^{\dag}_{2{\bf{k}},\uparrow}(\nu_{n})
\right]^{T}.
\label{Equation_A_3}
\end{eqnarray}
The indexes in Eq.(\ref{Equation_A_3}), before the wave vectors ${\bf{k}}$, mean the layer they are belonging to. For the layer-2 we get 
\begin{eqnarray}
\frac{\delta^{2}}{\delta{J^{\dag}_{2{\bf{k}},\uparrow}}\delta{J_{2{\bf{k}},\uparrow}}}=-\frac{1}{4}\left\langle c^{\dag}_{2{\bf{k}},\uparrow}(\nu_{n})c_{2{\bf{k}},\uparrow}(\nu_{n})\right\rangle=-\frac{1}{2}{G}^{33}_{{\bf{k}}}(\nu_{n}).
\label{Equation_A_4}
\end{eqnarray}
The function ${G}^{33}_{{\bf{k}}}(\nu_{n})$, is the component of the Green function matrix, inverse of the matrix $\hat{G}^{-1}_{{\bf{k}}}(\nu_{n})$. And for the average we get 
\begin{eqnarray}
G_{2,{\bf{k}}}(\nu_{n})=-\frac{-i\nu_{n}-\Delta_{c}+4t_{1}\gamma_{{\bf{k}}}+\bar{\mu}_{1}-\frac{U\bar{n}_{1}}{2}}{(i\nu_{n})^{2}+2i\nu_{n}a_{{\bf{k}}}+b_{{\bf{k}}}-(t_{\perp}+\Delta)^{2}}.
\label{Equation_A_5}
\end{eqnarray}
For the convenience, we have omitted the spin indexes in Eq.(\ref{Equation_A_2}).
The derivation of the Green function $G_{1,{\bf{k}}}(\nu_{n})$, for the layer-1, is analogue, and we have
\begin{eqnarray}
G_{1,{\bf{k}}}(\nu_{n})=-\frac{-i\nu_{n}-\Delta^{S}_{c}+4t_{2}\gamma_{{\bf{k}}}+\bar{\mu}_{2}-\frac{U\bar{n}_{2}}{2}}{(i\nu_{n})^{2}+2i\nu_{n}a_{{\bf{k}}}+b_{{\bf{k}}}-(t_{\perp}+\Delta)^{2}}.
\label{Equation_A_6}
\end{eqnarray}
The parameters $a_{{\bf{k}}}$ and $b_{{\bf{k}}}$, in Eqs.(\ref{Equation_A_5}) and (\ref{Equation_A_6}), are defined in Eqs.(\ref{Equation_28}) and (\ref{Equation_29}) in the Section \ref{sec:Section_5_1}.
With the conventions discussed in the Sections \ref{sec:Section_2} and \ref{sec:Section_3}, we have $\Delta^{S}_{c}=1/2$, and $\bar{n}_{l}=1$ for each layer. Furthermore, we can rewrite the Eqs.(\ref{Equation_A_5}) and (\ref{Equation_A_6}) in the more convenient forms:
\begin{eqnarray}
G_{2,{\bf{k}}}(\nu_{n})=-\frac{1}{2}\left(\frac{\eta_{{\bf{k}}}}{\sqrt{D_{{\bf{k}}}}}-1\right)\frac{1}{i\nu_{n}-\kappa_{1,{\bf{k}}}}+\frac{1}{2}\left(\frac{\eta_{{\bf{k}}}}{\sqrt{D_{{\bf{k}}}}}+1\right)\frac{1}{i\nu_{n}-\kappa_{2,{\bf{k}}}},
\label{Equation_A_7}
\end{eqnarray}
\begin{eqnarray}
G_{1,{\bf{k}}}(\nu_{n})=\frac{1}{2}\left(\frac{\eta_{{\bf{k}}}}{\sqrt{D_{{\bf{k}}}}}+1\right)\frac{1}{i\nu_{n}-\kappa_{1,{\bf{k}}}}-\frac{1}{2}\left(\frac{\eta_{{\bf{k}}}}{\sqrt{D_{{\bf{k}}}}}-1\right)\frac{1}{i\nu_{n}-\kappa_{2,{\bf{k}}}}.
\label{Equation_A_8}
\end{eqnarray}
The functions $\eta_{{\bf{k}}}$, $D_{{\bf{k}}}$ are given in Eqs.(\ref{Equation_33}) and (\ref{Equation_34}) in the Section \ref{sec:Section_5_1}. 
The parameters $\kappa_{i,{\bf{k}}}$ $i=1,2$, in the denominators of the functions in Eqs.(\ref{Equation_A_7}) and (\ref{Equation_A_8}) are defined in Eq.(\ref{Equation_24}), in the Section \ref{sec:Section_4}.

The single-particle spectral functions $A_{l,{\bf{k}}}(\nu)$ $l=1,2$, which are mostly demanded for the simple ARPES experimental evidence of the excitonic pair formation, and coherent condensate measurements (see the discussion in the Section \ref{sec:Section_5_1}), are defined as being the imaginary parts of the corresponding retarded Green functions \cite{cite-35}. Therefore, we write
\begin{eqnarray}
A_{l,{\bf{k}}}(\nu)=-\Im G_{l,{\bf{k}}}(\nu_{n})|_{i\nu_{n}\rightarrow \nu+i\alpha^{+}},
\label{Equation_A_9}
\end{eqnarray}
where the analytical continuation into the upper half complex semi-plane is done for the retarded function.
The single-particle DOS is then straightforward
\begin{eqnarray}
\rho_{l}(\nu)=\frac{1}{N}\sum_{{\bf{k}}}A_{l,{\bf{k}}}(\nu).
\label{Equation_A_10}
\end{eqnarray}
Next, we will use the Cauchy's relation $1/(x+i\alpha^{+})\rightarrow {\cal{P}}(1/x)-i\pi\delta(x)$ in order to separate the real and imaginary parts, in the expressions in Eqs.(\ref{Equation_A_7}) and (\ref{Equation_A_8}). We get
\begin{eqnarray}
\rho_{l}(\nu)=\sum_{i=1,2}\frac{(-1)^{l+i-2}}{2N}\sum_{{\bf{k}}}\left[\frac{\eta_{{\bf{k}}}}{\sqrt{D_{{\bf{k}}}}}-(-1)^{l+i-1}\right]\delta\left(\nu-\kappa_{i,{\bf{k}}}\right).
\label{Equation_A_11}
\end{eqnarray}
The summation over the 2D wave vectors ${\bf{k}}$ in Eq.(\ref{Equation_A_11}) could be transformed into the integration, by introducing the density of states function $\rho_{\rm 2D}(z)=\frac{1}{N}\sum_{{\bf{k}}}\delta(z-\gamma_{{\bf{k}}})$, where $\gamma_{{\bf{k}}}$ is the 2D lattice dispersion and is given explicitly in Eq.(\ref{Equation_20}) in the Section \ref{sec:Section_5_2} . 
Then, the Eq.(\ref{Equation_A_11}) will be transformed into following equation
\begin{eqnarray}
\rho_{l}(\nu)=\sum_{i=1,2}\frac{(-1)^{l+i-2}}{2}\int{dy}\rho(y)\left[\frac{\eta(y)}{\sqrt{D(y)}}-(-1)^{l+i-1}\right]\delta\left[\nu-\kappa_{i}(y)\right].
\label{Equation_A_12}
\end{eqnarray}
We can now invert the Dirac $\delta$ function in Eq.(\ref{Equation_A_12}), in order to perform the integration over the continuous variable of integration. This is done easily, by using the relation for the Dirac function: $\delta\left[f(x)\right]=\sum_{i}{\delta(x-x_{i})}/{|f'(x_{i})|}$,
where $x_{i}$ are the solutions of the equation $f(x)=0$. 
Then, we obtain the following compact formula for the $l$-layer single-particle excitonic DOS function:
\begin{eqnarray}
\rho_{l}(\nu)=\sum_{\substack{i=1,2\\ j=1,2}}\frac{(-1)^{l+j-2}\rho_{\rm 2D}(\epsilon_{i})}{2|\Lambda_{j}(\epsilon_{i})|}\left[\frac{\eta(\epsilon_{i})}{\sqrt{D(\epsilon_{i})}}-(-1)^{l+j-1}\right],
\label{Equation_A_13}
\end{eqnarray}
where the paremeters $\Lambda_{j}(z)$ $j=1,2$ are given in Eq.(\ref{Equation_44}) in the Section \ref{sec:Section_5_2}. The
parameters $\epsilon_{i}$ i = 1,2, in the arguments of the functions
in Eq.(\ref{Equation_A_13}), are also defined in the Section \ref{sec:Section_5_2} in Eq.(\ref{Equation_45}).

The numerical evaluations of the DOS functions, obtained with Eq.(\ref{Equation_A_13}), for the layers $l=1,2$, and for different values of the interlayer interaction parameter $W/|t_{1}|$, are given in the Section \ref{sec:Section_5_2} of the present paper. 
%
\section*{References}
%

%
\end{document}